\documentclass[twocolumn]{aastex63}

\newcommand{\mdiff}{\mathrm{d}}
\newcommand{\amin}{$^{\prime}$}
\newcommand{\asec}{$^{\prime\prime}$}
\newcommand{\kms}{\,km\,s$^{-1}$}
\newcommand{\um}{\,$\mu$m}
 
\newcommand{\cp} {C$^{+}$}

\newcommand{\oi} {{\sc [O\,i]}}

\newcommand{\cii}{{\sc [C\,ii]}} 
\newcommand{\nii}{{\sc [N\,ii]}} 
 
\newcommand{\hi}{{\sc H\,i}} 
\newcommand{\hii}{{\sc H\,ii}} 
 
\newcommand{\oisw}{\mbox{{\sc [O\,i]}\,63\,$\mu$m}}

\newcommand{\ciiw}{\mbox{{\sc [C\,ii]}\,158\,$\mu$m}}

\newcommand{\thco}{$^{13}$C$^{}$O}
\newcommand{\ceio}{$^{}$C$^{18}$O}

\newcommand {\ra}[3]{$\alpha_{J2000} = #1^{\rm h} #2^{\rm m} #3^{\rm s}$} 
\newcommand {\dec}[3]{$\delta_{J2000} = #1^\circ #2^\prime #3^{\prime\prime}$} 

\newcommand \percmsq {\,cm$^{-2}$}
\newcommand \percmcu {\,cm$^{-3}$}

\newcommand \msun {\,M$_\odot$}
\newcommand \lsun {\,L$_\odot$}

\newcommand \sgra {Sgr\,A}
\newcommand \sgrb {Sgr\,B}
\newcommand \sgrbo {Sgr\,B1}
\newcommand \sgrbt {Sgr\,B2}
\newcommand \grt {upGREAT}
\newcommand \gsix {G\,0.6--0.0}

\newcommand \pell {$+\ell$}

\usepackage{enumitem}
\begin{document}

\title{SOFIA-\grt\ imaging spectroscopy of the \ciiw\ fine structure
  line \\of the Sgr\,B region in the Galactic center}

\received{2021 May 31}
\revised{2021 July 26}
\shorttitle{\cii\ toward \sgrb}
\shortauthors{Harris et al.}
\correspondingauthor{Andrew Harris}
\email{harris@astro.umd.edu}

\author{A.I.~Harris} 
\affil{Department of Astronomy, University of Maryland, College Park, MD 20742, USA}
\author{R.~G\"usten} 
\affiliation{Max-Planck-Institut f\"ur Radioastronomie, Auf dem H\"ugel 69, 53121 Bonn, Germany}
\author{M.A.~Requena-Torres} 
\affiliation{Department of Astronomy, University of Maryland, College Park, MD 20742, USA}
\affiliation{Department of Physics, Astronomy, and Geosciences, Towson University, Towson, MD 21252, USA}
\author{D.~Riquelme}
\affiliation{Max-Planck-Institut f\"ur Radioastronomie, Auf dem
  H\"ugel 69, 53121 Bonn, Germany}
\author{M.R.~Morris} 
\affiliation{Department of Physics and Astronomy, University of California, Los Angeles, CA 90095, USA}
\author{G.J.~Stacey} 
\affiliation{Department of Astronomy, Cornell University, Ithaca, NY
  14853, USA}
\author{J.~Mart\`in-Pintado} 
\affiliation{Centro de Astrobiolog\'ia (CSIC-INTA), Ctra.\ de Ajalvir
  Km.\ 4, 28850, Torrej\`on de Ardoz, Madrid, Spain}
\author{J.~Stutzki} 
\affiliation{I. Physikalisches Institut der Universit\"at zu K\"oln, Z\"ulpicher Stra{\ss}e 77, 50937 K\"oln, Germany}
\author{R.~Simon} 
\affiliation{I. Physikalisches Institut der Universit\"at zu K\"oln, Z\"ulpicher Stra{\ss}e 77, 50937 K\"oln, Germany}
\author{R.~Higgins} 
\affiliation{I. Physikalisches Institut der Universit\"at zu K\"oln, Z\"ulpicher Stra{\ss}e 77, 50937 K\"oln, Germany}
\author{C.~Risacher} 
\affiliation{Max-Planck-Institut f\"ur Radioastronomie, Auf dem
  H\"ugel 69, 53121 Bonn, Germany}
\affiliation{now Institut de Radioastronomie Millim\`{e}trique, 300 rue de
  la Piscine, Domaine Universitaire, 38406 Saint Martin d'H\`{e}res, France}

\begin{abstract}
We report SOFIA-\grt\ spectroscopic imaging of the \ciiw\ spectral
line, as well as a number of \oisw\ spectra, across a $67 \times
45$\,pc field toward the \sgrb\ region in our Galactic center.  The
fully-sampled and velocity-resolved \cii\ images have 0.55\,pc spatial
and 1\kms\ velocity resolutions.

We find that \sgrb\ extends as a coherent structure spanning some
34\,pc along the Galactic plane.  Bright \cii\ emission encompasses
\sgrbo\ (G0.5--0.0), the \gsix\ \hii\ region, and passes behind and
beyond the luminous star forming cores toward \sgrbt\ (G0.7--0.0).
\sgrb\ is a major contributor to the entire Galactic center's \cii\
luminosity, with surface brightness comparable to \cii\ from the
Arches region.

\cii, 70\um, and 20\,cm emission share nearly identical spatial
distributions.  Combined with the lack of \cii\ self-absorption, this
indicates that these probes trace UV on the near surfaces of more
extended clouds visible in CO isotopologues and 160\um\ continuum.
Stars from regions of local star formation likely dominate the UV
field. Photodissociation regions and \hii\ regions contribute similar
amounts of \cii\ flux.

The extreme star formation cores of \sgrbt\ contribute negligible
amounts to the total \cii\ intensity from the \sgrb\ region. Velocity
fields and association with a narrow dust lane indicate that they may
have been produced in a local cloud-cloud collision.  The cores are
likely local analogs of the intense star formation regions where ideas
to explain the ``\cp\ deficit'' in ultra-luminous galaxies can be
tested.
  
\end{abstract}



\section{Introduction} \label{sec:intro}

We present spectrally resolved imaging of the \ciiw\ spectral line
from the luminous \sgrb\ region in our Galactic center. Understanding
the distribution of \cii\ intensity and its relationship with star
formation in galactic nuclei is important for understanding our
galaxy, as well as for interpreting observations of galaxies
throughout the universe.  Velocity-resolved spectroscopy allows
separation of different physical components and investigation of
dynamical and physical states of the interstellar medium, and
structural localization at the sub-beam level.

The \cii\ fine structure line from singly-ionized carbon is a powerful
probe of the interstellar medium.  A major cooling line, it is often
the brightest spectral indicator of star formation within our Galaxy
and in others \citep[e.g.,][and references therein]{crawford85,
  stacey91, boselli02, delooze14, pineda14, herreracamus15}. Even at
energies below the hydrogen ionization edge, soft ultraviolet
radiation ionizes atomic carbon in a range of environments: neutral
diffuse material, the warm ionized interstellar medium, and warm and
dense molecular gas.  Collisions with electrons, atomic hydrogen, and
molecular hydrogen populate the transition's upper
level. Photochemistry models (e.g., \citealt{tielens85, vandishoeck88,
  wolfire90, sternberg95, kaufman99}) support interpretation of line
intensities as functions of physical conditions.  Extreme conditions
in ultra-luminous galactic nuclei (ULIRGs) suppress \cii\ emission,
however (\citealt{fischer99} and references therein).

\sgrb\ is one of the most luminous regions within the Galaxy's Central
Molecular Zone (CMZ), a region of particularly dense, warm, and
turbulent molecular clouds in the inner few hundred parsecs of the
Galaxy (reviews by \citealt{guesten86} and \citealt{morris96} contain
overviews of the CMZ and its basic properties; \citealt{mills17}
provides recent updates).  \sgrb\ contains extended emission regions
and high-density stellar clusters embedded in dense envelopes, making
it an excellent model for close examination of star formation within
galactic nuclei.  Its proximity (we use 8\,kpc; \citealt{reid14} find
the distance to \sgrbt\ is 7.9\,kpc, and \sgrbt\ appears to be in
front of the main \sgrb\ cloud) allows detailed studies at many
wavelengths. The \sgrb\ region and some of its components were first
studied in single-dish radio continuum mapping of the Galactic center
\citep{kapitzky74, mezger76, downes80}.  These observations identified
three main sources: G\,0.5--0.0 (\sgrbo), a bright peak \gsix, and a
yet brighter region G\,0.7--0.0 (\sgrbt).  The designations were based
simply on radio continuum brightness peaks, and did not necessarily
identify physically isolated or connected regions.  Subsequent
observations at higher spatial resolutions showed that the individual
sources contain multiple components.  Interferometric radio continuum
imaging \citep{mehringer92, lang10} showed arcs and bars of emission
across \sgrbo, compact sources in \gsix, and very bright compact cores
within \sgrbt.  In this paper we associate the \gsix\ and \sgrbo\
regions with extended \cii-bright emission that surrounds those
sources and extends further to positive latitude, since these all
appear to be parts of the same general structure, while the \sgrbt\
cores \sgrbt(M) and (N), and perhaps (S), seem to be quite distinct
foreground objects.

While the brightest material surrounding the dense young stellar
clusters in \sgrbt's cores have been of great interest for many years,
the larger region encompassing \sgrbt, \sgrbo, \gsix, and their
diffuse envelope emission, has received much less attention.
\citet{goi04} identified extended far-IR fine structure line emission
around \sgrbo\ and \sgrbt\ in their {\em ISO} satellite cross-cuts of
the region with the LWS grating spectrometer.  \citet{simpson18b} and
\citet{simpson21} imaged the \sgrbo\ region in mid-and far-IR lines to
investigate conditions in the ionized and neutral material.
\citet{santamaria21} used the {\em Herschel}-SPIRE FTS to make a broad
spectral survey of the region around the luminous \sgrbt\ star
formation cores.  None of these observations had sufficient spectral
resolution to separate components within the beams by velocity,
however.

Single pointings toward the most luminous core, \sgrbt(M), found \cii\
and \oisw\ absorption from material from the Galactic plane in {\em
  ISO} LWS Fabry-Perot observations \citep{goi04}, and {\em
  Herschel}-HIFI made high spectral resolution observations in a much
smaller beam as part of the HEXOS key project that showed absorption
near the center of the emission line \citep{bergin10, moeller21}.
\citet{langer17} observed \cii\ emission in strip maps toward the edge
of the CMZ with {\em Herschel}-HIFI.  While their observations had the
high spectral resolution needed to identify different physical
components, their strips close to \sgrb\ did not cross any of its main
clouds.

Here we report full two-dimensional spectroscopic imaging of the
\sgrb\ complex with 0.55\,pc and 1\kms\
resolutions. Section~\ref{sec:obs} describes our observations,
Sec.~\ref{sec:results} is a summary of the main results, and
Sec.~\ref{sec:discus} discusses implications for the Galactic center
and interpretation of observations of external galaxies.

\begin{figure}[!ht]
\centering
\includegraphics[width=0.45\textwidth]
{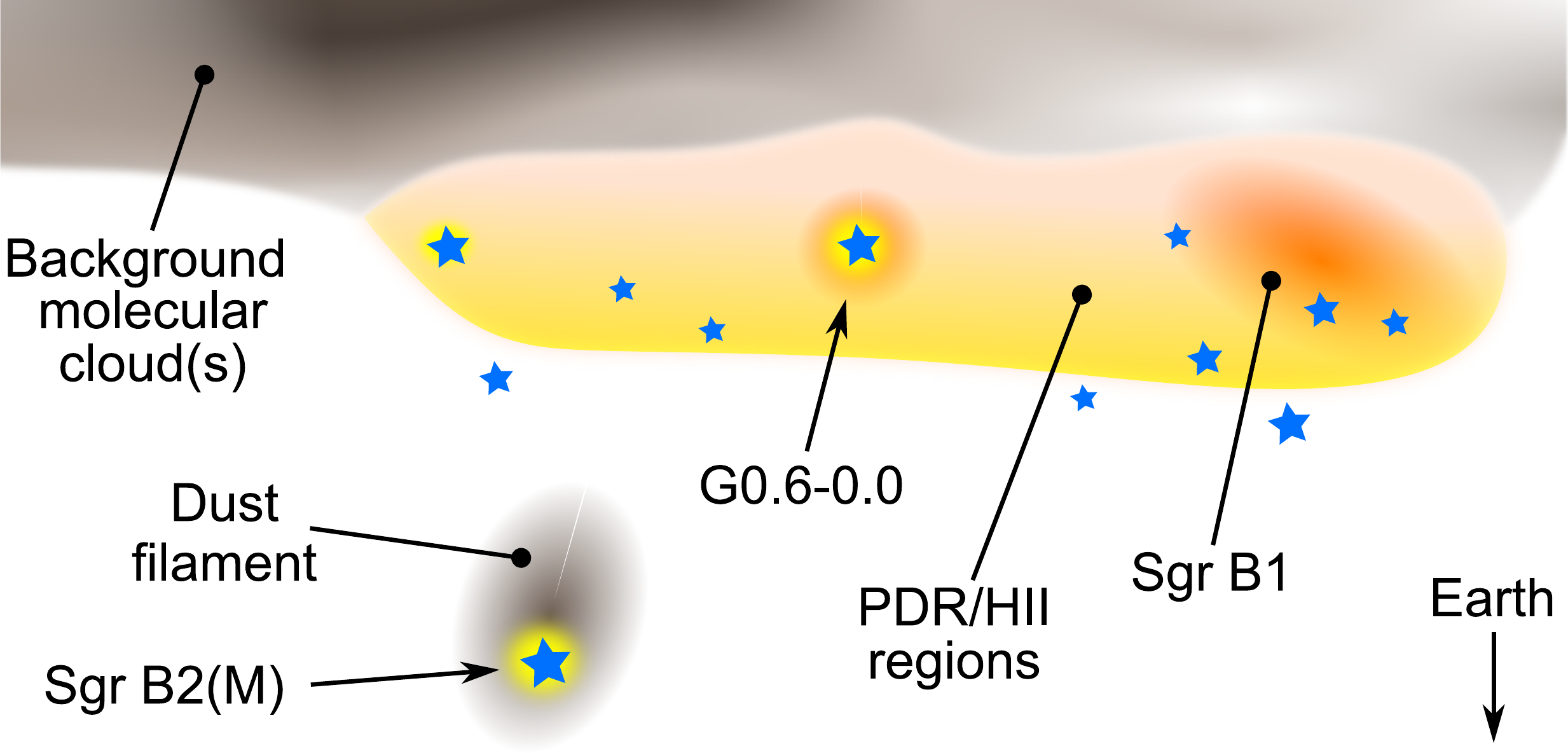}
\caption{Schematic ``top view'' sketch of the \sgrb\ region showing
the relative positions of the region's components, to no particular
scale, cut by a plane notionally defined by the Earth and the
position-velocity cut line in Fig.~\ref{fig:sgrbregion}.  The
separation between the dust filament and the background cloud is
unknown, as emphasized by a gap in the sketch, although the velocity
of the filament and cloud are close.  Our perspective is from the
bottom of the page.
  \label{fig:sketch}}
\end{figure}

To set a preliminary framework for discussion, Fig.~\ref{fig:sketch}
is a sketch of the physical structure of the region.  We explore our
logic for reaching this view in Sec.~\ref{ssec:physstruct}, but in
brief: \cii\ velocities and its spatial distribution indicate that the
\sgrb\ region is a coherent structure that physically incorporates
\sgrbo\ and \gsix, then continues past the Galactic latitude of the
\sgrbt\ cores.  The bright \sgrbt\ star formation cores appear to be
associated with the dark dust lane that crosses in front (closer to
Earth) of the main \sgrb\ cloud, which may or may not be physically
connected to the main \cii-emitting region.  Lack of \cii\
self-absorption across the \sgrb\ body places the \cii\ and \oi\ at or
close to the surface of a larger background (further from Earth) cloud
or clouds visible in molecular and long-wavelength dust emission.
Loosely-distributed stars in and near the surface provide UV photons
to ionize, excite, and heat the gas and dust in the surface layer.

Sec.~\ref{sec:summary} provides brief summaries of our main findings.

Far-infrared observations have ties to both infrared and radio
conventions, so we quote intensities $I$ in units either energy-based
(e.g., $[\mathrm{erg~s^{-1}~cm^{-2}~sr^{-1}}]$) or velocity-integrated
brightness temperatures (e.g., [K\kms]) depending on context.  The two
are related by
\begin{equation}
I = \int I_\nu \, \mdiff\nu = \frac{2k}{\lambda^3} 
\int T_B \, \mdiff{\rm v}
\label{eq:intens}
\end{equation}
through the Rayleigh-Jeans expansion that defines brightness
temperature $T_B$ from specific intensity $I_\nu$, and the Doppler
relationship between frequency and velocity.

\begin{figure*}[!ht]
\centering
\includegraphics[width=\textwidth]
{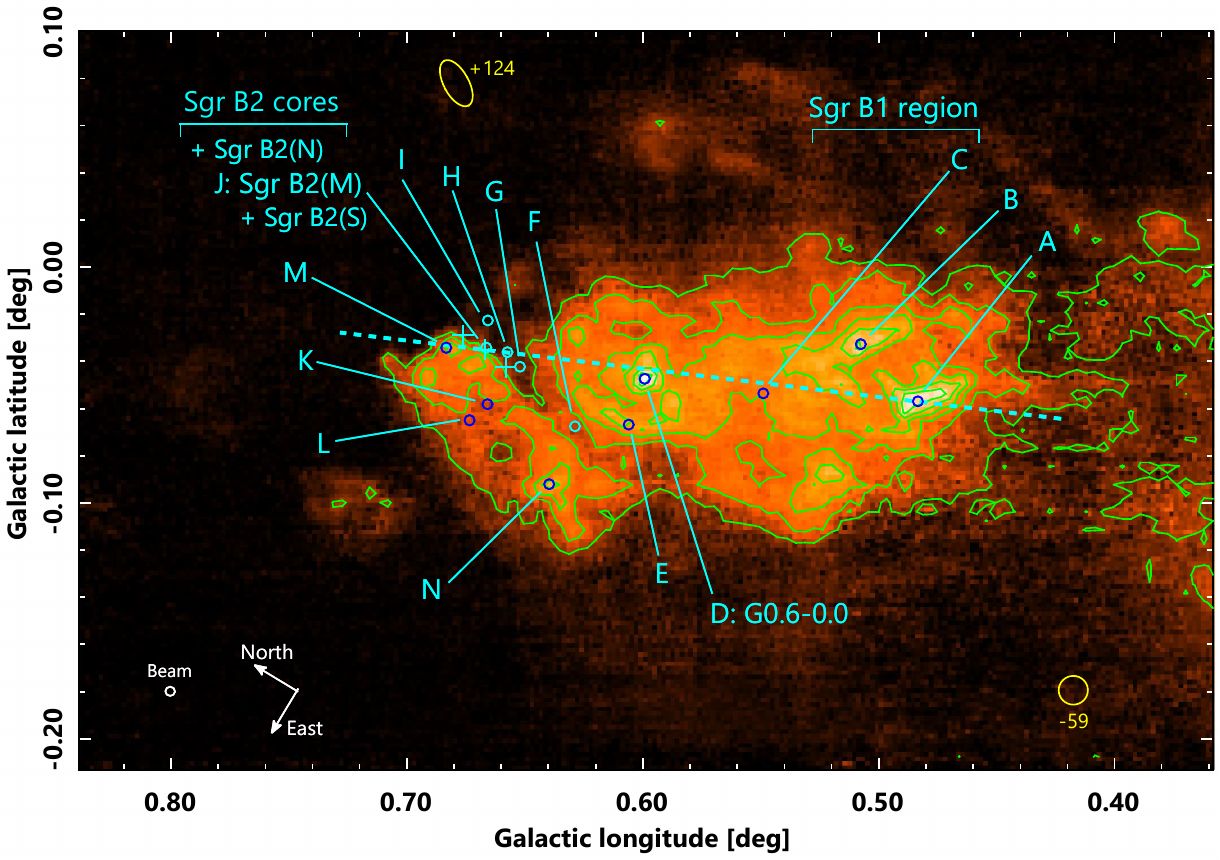}
\caption{
\ciiw\ integrated intensity image of the \sgrb\ region covering
14--120\kms.  The image covers $\Delta \ell = 0.48\degr$ in Galactic
longitude and $\Delta b = 0.32\degr$ in longitude, centered at $\ell =
0.598\degr$, $b = -0.057\degr$.  For a distance of 8\,kpc, these
angles correspond to $67.0 \times 44.7$\,pc, sampled in a 0.55\,pc
beam.  The main \cii\ emission region is about 0.24\degr\ long and
0.11\degr\ wide (34\,pc$\times 15$\,pc).  Contour levels show
intensities of 175 to 550\,K\kms\ in steps of 75\,K\kms\ within
14.1\asec\ FWHM beams.  Circles with letter labels denote positions
for the spectra in Fig.~\ref{fig:spectralstack}; crosses mark the
positions of \sgrbt(N), (M), and (S).  The dotted straight line shows
the path of the position-velocity cut in Fig.~\ref{fig:pv}.  The
yellow ellipse near the top of the image and circle to the bottom
right show the extents of compact emission sources at $+124$ and
$-59$\kms\ (Sec.~\ref{sec:compact}).
\label{fig:sgrbregion}}
\end{figure*}

\section{Observations} \label{sec:obs}

Our results are part of our larger \cii\ spectral data cube
\citep{wholecmz} of the Central Molecular Zone.  Our larger cube
covers $1.5\degr$ in longitude and $0.32\degr$ in latitude.  The map
center for this entire CMZ project was \ra{17}{47}{24.60},
\dec{-28}{23}{16.3}.  Data presented here were obtained during two
observing campaigns flying from New Zealand in 2017 June and July and
2018 June, yielding the most positive Galactic longitude ($\ell$)
$3 \times 2$ ``tiles'' that formed the basic structure of the imaging
project.  Each tile covered $560 \times 560$ arcseconds.  Together,
these tiles covered an area $\Delta\ell = 0.48\degr$ by
$\Delta b = 0.32\degr$ centered at $\ell = 0.598\degr$,
$b = -0.057\degr$.

Each tile was mapped in total power on-the-fly mode in rotated Right
Ascension-Declination frames corresponding to Galactic coordinates.
Individual integrations were recorded every 0.3\,sec, with spacing of
7\asec\ along the scan direction; final images were convolved to
15\asec\ FWHM Gaussian beams on a 6.9\asec\ rectangular grid.  The rms
pointing accuracy was 2\asec.  Tile offset positions allowed for one
row (column) of overlap between neighboring tiles.  Each tile was
observed in orthogonal directions to reduce striping and residual
scanning structure.  We used two reference ``off'' positions: a
relatively nearby position with some expected line contamination
during mapping (\ra{17}{47}{41.3}, \dec{-28}{35}{00}), and a distant
field far from the Galactic plane (\ra{17}{55}{03.9},
\dec{29}{23}{02}) to measure and correct the contamination in the
closer ``off'' position.

We observed the 1.901\,THz ($\lambda$157.74\um) \cii\
$^2P_{3/2} - ^2P_{1/2}$ and the 4.748\,THz ($\lambda$63.18\um) \oi\
\mbox{ } $^{3}P_{1} - ^{3}P_{2}$ fine structure transitions with the
upgraded German Receiver for Astronomy at Terahertz Frequencies
(\grt\footnote{\grt\ is a development by the Max-Planck-Institut f\"ur
  Radioastronomie and the I.~Physikalisches Institut of the
  Universit\"at zu K\"oln, in cooperation with the DLR Institut f\"ur
  Optische Sensorsysteme.}) \citep{risacher18} on the Stratospheric
Observatory For Infrared Astronomy (SOFIA, \citealt{young12}).  Rapid
imaging of \cii, with auxiliary data in \oi\ (the observing strategy
was geared to the brighter \cii\ line), was possible with \grt's dual
frequency focal plane array configured for parallel observations with
both the low frequency arrays (LFA) and high frequency array (HFA).
Main beam sizes were 14.1\asec\ for the LFA, and 6.3\asec\ for the
HFA. Both arrays use niobium nitride hot electron bolometer mixers,
pumped with either a solid-state source (LFA) or quantum cascade laser
(HFA) local oscillator.  The LFA has seven dual-polarized pixels in a
hexagonal arrangement with 31.8\asec\ radial spacing around a central
pixel. The HFA has the same symmetry, centered on the LFA's central
pixel, with single-polarized pixels on 13.8\asec\ radial spacing. Fast
Fourier Transform Spectrometers (FFTS4G, updated from
\citealt{klein12}) produced spectra across the 0-4\,GHz intermediate
frequency bands, with 32k channels binned for 1\kms\ velocity
resolution.

Raw data were amplitude calibrated with the $kosma\_kalibrate$
software package (versions 2017.08 and 2018.07) following
\citet{guan12}.  Line temperatures are on a $T_{\rm MB}$ scale, with
estimated absolute uncertainty below 20\%.  Small but repeatable
receiver instabilities produced spectral baseline structure that we
removed using a family of baseline structures derived from differences
between nearby ``off'' spectra \citep{higgins11,
  kester14, higgins21}. This method retains information in broad lines
and is superior to partly subjective low-order polynomial fits.
Further processing used the GILDAS packages CLASS and GREG.

\section{Results}\label{sec:results}

\subsection{\cii\ emission from the main body of the \sgrb\ cloud \label{ssec:ciispect}}

\begin{figure}[!ht]
\centering
\includegraphics[width=0.45\textwidth]{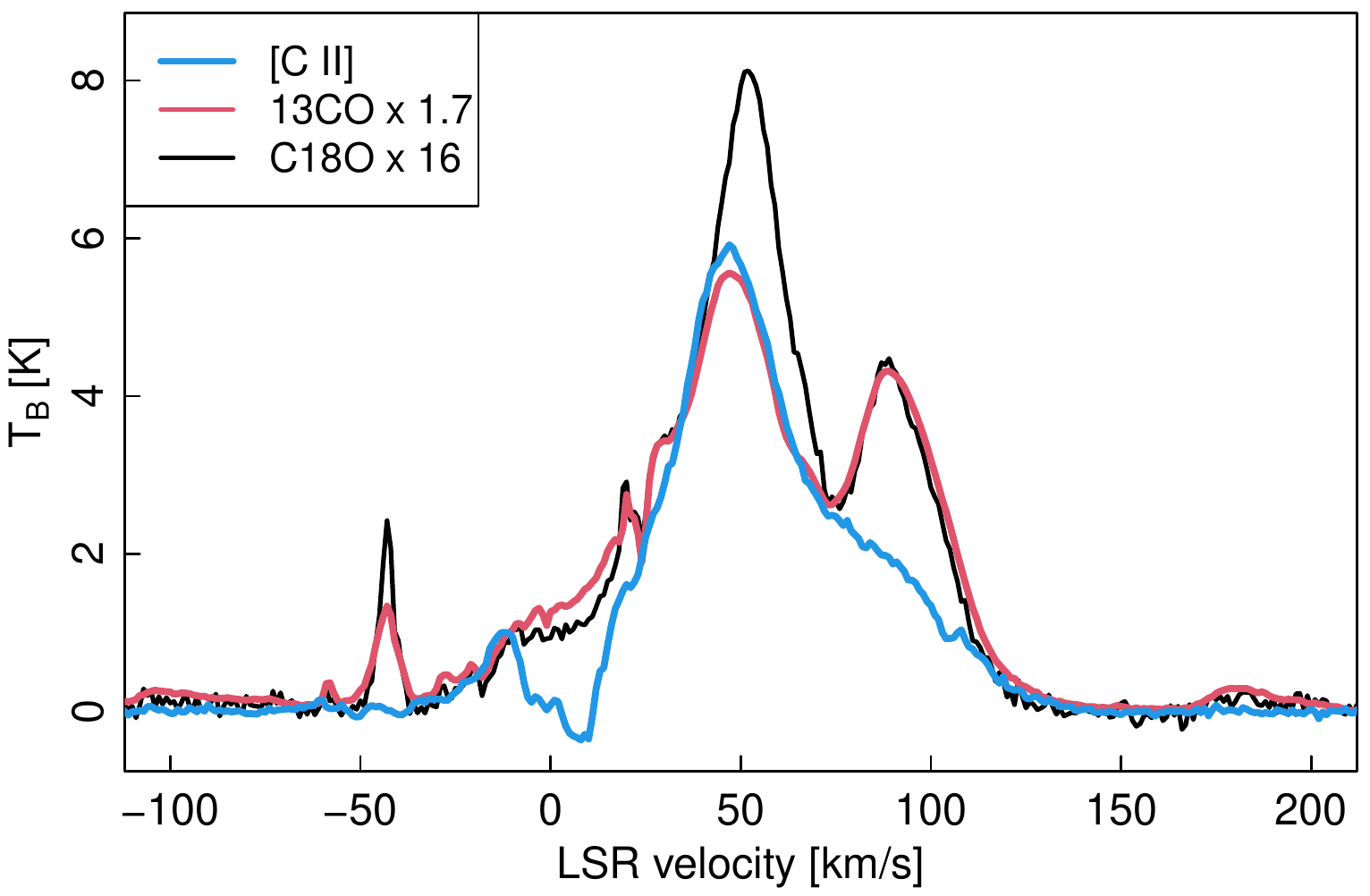} 
\caption{ Lineshape comparisons between \cii, \thco\ $J=2-1$, and \ceio\
  $J=2-1$ \citep{apexco} emission averaged over the \sgrb\ region
  indicated by the lowest (175\,K\kms) contour in
  Fig.~\ref{fig:sgrbregion}. Line of sight \cii\ absorption, or
  emission in the reference position,  accounts
  for the lack of flux near 0\kms. \label{fig:ciico}}
\end{figure}

\begin{figure*}[!ht]
\centering
\includegraphics[width=\textwidth]{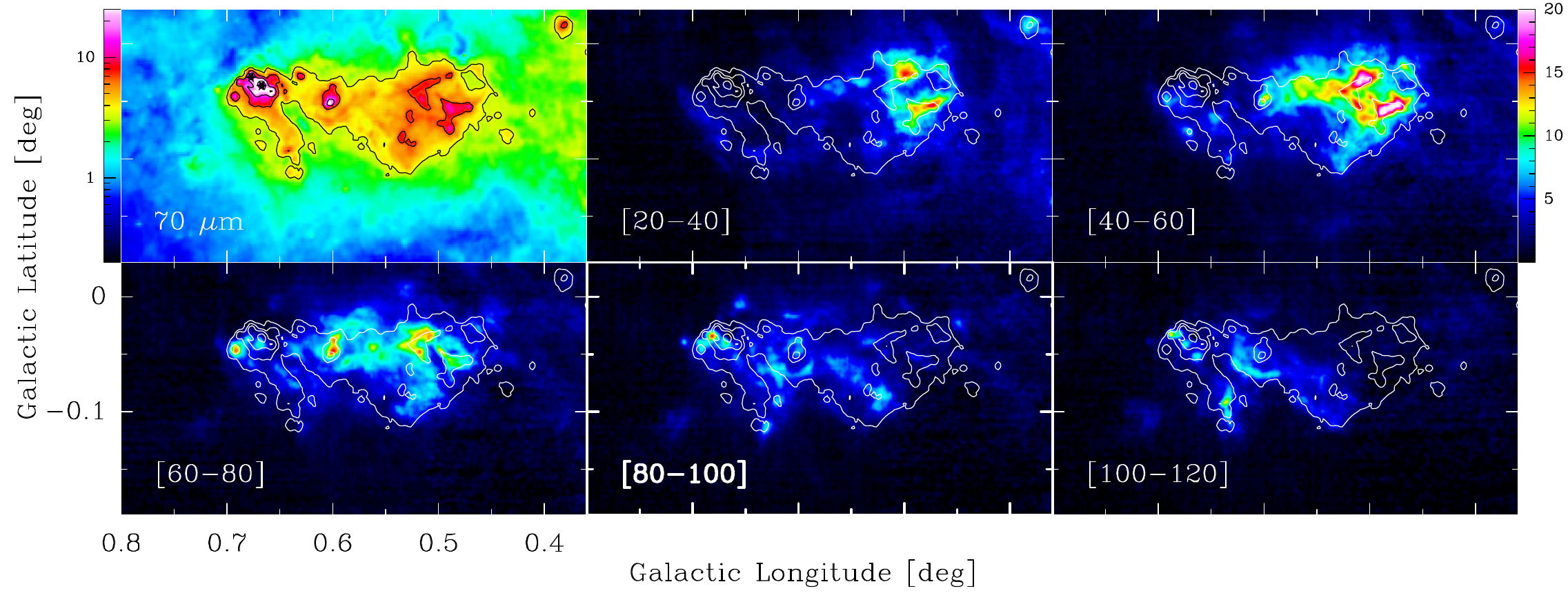}
\caption{ \cii\ channel maps as images, in 20\kms\ wide velocity
  bins.  The wedge to the right defines the respective colors between 0 and 20
K km/s.  Contours are 70\um\ continuum levels from \citet{molinari16}
at 2, 4, 10 and 40\% of the field's peak intensity at \sgrbt(M). A
false-color image of the latter is in the top left panel, on a log
scale to emphasize lower level emission, with positions of \sgrbt(M)
and (N) marked by black stars.
\label{fig:chanmaps}}
\end{figure*}

\begin{figure*}[!ht]
\centering
\includegraphics[width=0.7\textwidth]{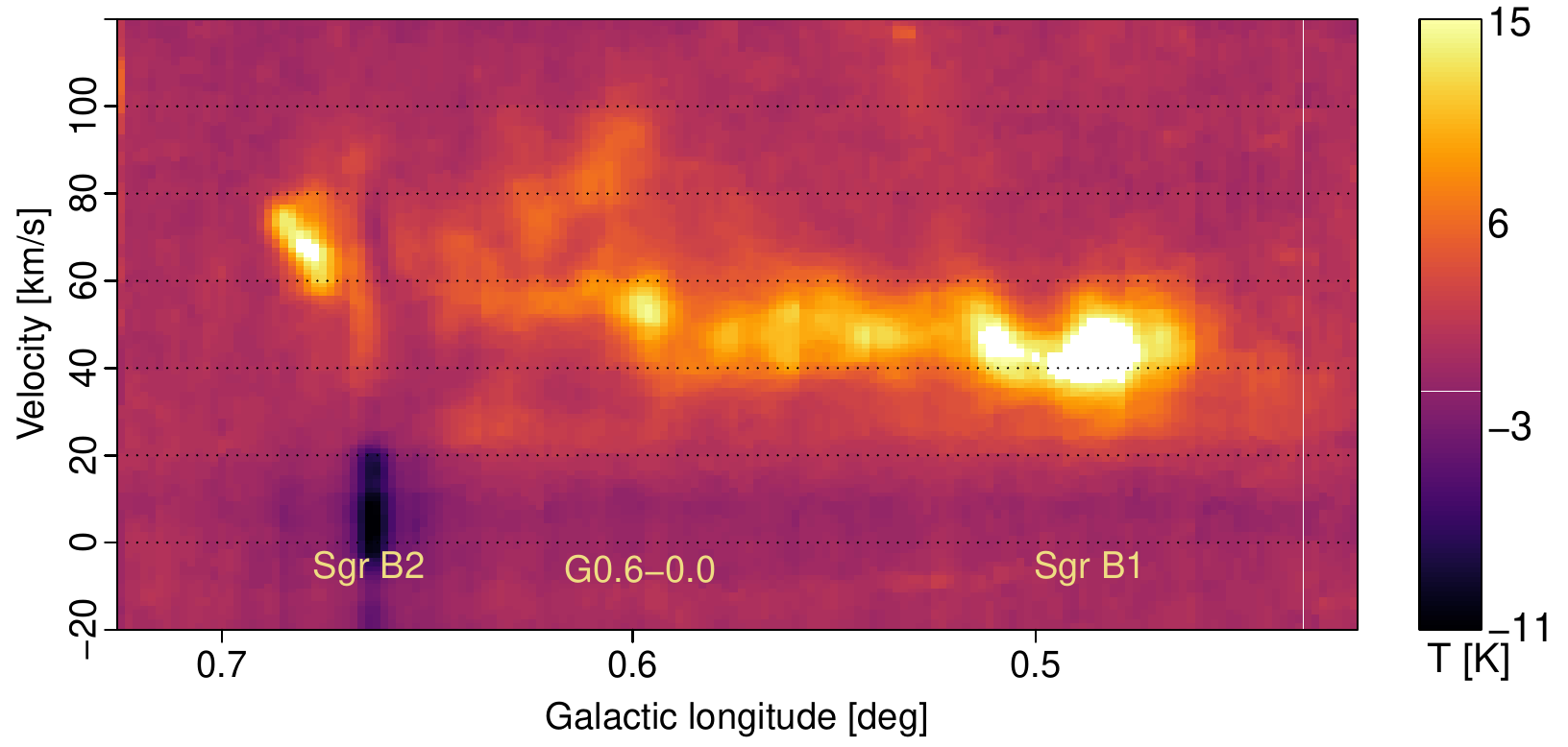}
\caption{ Position-velocity cut along a line marked by the dotted line
  in Fig.~\ref{fig:sgrbregion}, with positions projected onto Galactic
  longitude.  The cut passes through the \cii\ peak at region A, near
  region D at \gsix, and continues through \sgrbt(M) at region J, and
  past the peak at region M.  Brightness temperatures above 15\,K
  saturate at the lightest color in this plot.  \cp\ in the Galactic
  plane along the line of sight to the \sgrbt(M) continuum source
  accounts for the deep absorption at $\ell \sim 0.67\degr$ across 0
  to 20\kms. Absorption shows as negative because \sgrbt(M)'s
  continuum offset was removed in spectral baseline fitting.
\label{fig:pv}}
\end{figure*}

\begin{figure*}[t]
\centering
\includegraphics[width=0.7\textwidth]{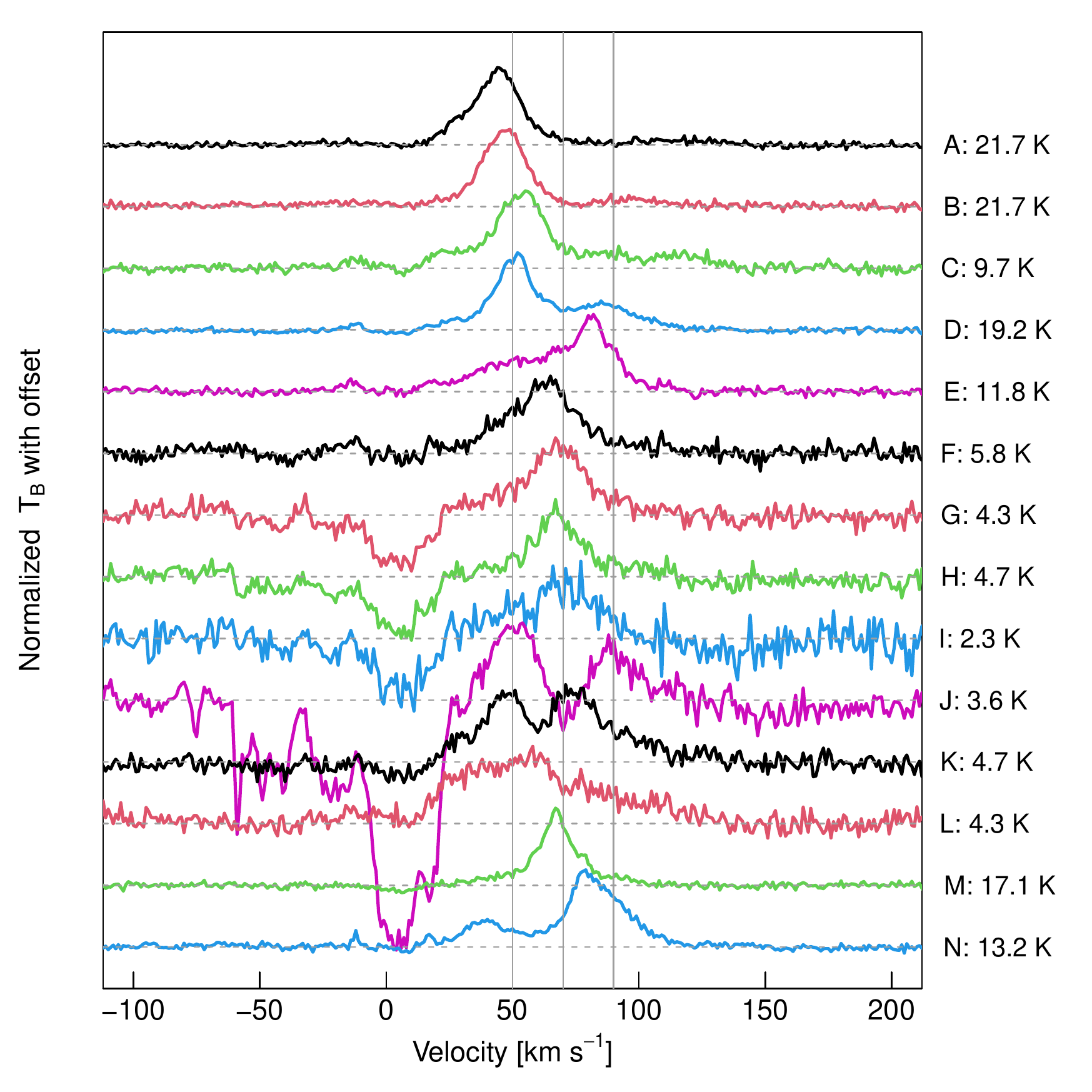}
\caption{ \cii\ spectra in single beams from the positions and areas
  marked in Fig.~\ref{fig:sgrbregion}. The spectra are normalized to
  unity by peak temperatures, which are given after each label.  Light
  vertical lines mark 50, 70, and 90\kms\ to assist in comparing velocity
  shifts.  Parameters for two-component Gaussian fits to emission are
  in Table~\ref{tab:fits}.
  \label{fig:spectralstack}}
\end{figure*}

\begin{deluxetable*}{crrrrrrrrrr}[t]
\tabletypesize{\scriptsize}
\tablecaption{Two-component Gaussian fit results to the main
  components of the 
  spectra of Fig.~\ref{fig:spectralstack}.  Letters refer to positions
  in Fig.~\ref{fig:sgrbregion}.  Line parameter columns give main-beam
  brightness temperature, line center, line width, and intensity of
  the two main components.  The component with the larger peak
  brightness temperature is listed first; dashes show that a second
  very broad component was likely residual baseline
  structure. Horizontal lines represent some separation by region.
  \sgrbt(M) is at position J.
  \label{tab:fits}}
  \tablewidth{0pt}
  \tablecolumns{9}
  \tablehead{ \colhead{Pos.} & 
  \colhead{$\ell$} &
  \colhead{$b$} &
  \colhead{$T_1$} &
  \colhead{$v_1$} &
  \colhead{$\Delta v_1$} &
  \colhead{$I_1$} &
  \colhead{$T_{2}$} &
  \colhead{$v_2$} &
  \colhead{$\Delta v_2$} &
  \colhead{$I_2$} \vspace{-3mm} \\ 
  &
  \colhead{[deg]} &
  \colhead{[deg]} &
  \colhead{[K]} &
  \colhead{[\kms]} &
  \colhead{[\kms]} &
  \colhead{[K\kms]} &
  \colhead{[K]} &
  \colhead{[\kms]} &
  \colhead{[\kms]} &
  \colhead{[K\kms]} }
\startdata
A & 0.483 & -0.057 & $13.4 \pm 0.9$ & $42.2 \pm 0.4$ & $49.3 \pm 1.6$ & $701 \pm 54$ & $8.5 \pm 0.9$ & $45.7 \pm 0.3$ & $19.7 \pm 1.8$ & $177 \pm 25$ \\ 
B & 0.507 & -0.033 & $20.1 \pm 0.3$ & $46.6 \pm 0.1$ & $32.6 \pm 0.5$ & $699 \pm 14$ & --- & --- & --- & --- \\
C & 0.549 & -0.054 & $7.6 \pm 0.2$ & $53.1 \pm 0.2$ & $31.7 \pm 1.1$ & $257 \pm 12$ & --- & --- & --- & --- \\
\hline
D & 0.599 & -0.047 & $16.8 \pm 0.3$ & $50.3 \pm 0.2$ & $32.8 \pm 0.8$ & $587 \pm 17$ & $6.3 \pm 0.2$ & $84.4 \pm 0.8$ & $57.0 \pm 3.2$ & $380 \pm 25$ \\ 
E & 0.606 & -0.067 & $7.2 \pm 0.3$ & $82.3 \pm 0.2$ & $25.6 \pm 1.1$ & $195 \pm 11$ & $5.3 \pm 0.1$ & $61.8 \pm 0.9$ & $93.4 \pm 2.4$ & $523 \pm 20$ \\ 
\hline
F & 0.629 & -0.068 & $3.6 \pm 0.3$ & $61.2 \pm 0.7$ & $69.8 \pm 3.8$ & $265 \pm 28$ & $1.9 \pm 0.3$ & $64.2 \pm 0.7$ & $21.2 \pm 4.0$ & $43 \pm 11$ \\ 
G & 0.652 & -0.043 & $2.3 \pm 0.9$ & $65.7 \pm 2.4$ & $67.3 \pm 13.2$ & $166 \pm 72$ & $1.7 \pm 0.9$ & $68.8 \pm 1.8$ & $24.6 \pm 11.7$ & $43 \pm 31$ \\ 
H & 0.657 & -0.036 & $2.1 \pm 0.9$ & $66.8 \pm 1.5$ & $21.1 \pm 9.2$ & $47 \pm 29$ & $1.8 \pm 0.9$ & $69.0 \pm 3.3$ & $65.5 \pm 18.5$ & $126 \pm 72$ \\ 
I & 0.665 & -0.023 & $1.5 \pm 0.2$ & $70.8 \pm 2.9$ & $49.4 \pm 11.0$ & $80 \pm 20$ & $0.6 \pm 0.3$ & $42.6 \pm 5.4$ & $32.1 \pm 18.4$ & $20 \pm 15$ \\ 
J & 0.666 & -0.034 & $7.5 \pm 2.2$ & $67.8 \pm 0.7$ & $66.4 \pm 5.6$ & $530 \pm 160$ & $-8.2 \pm 2.1$ & $70.0 \pm 0.3$ & $37.3 \pm 3.5$ & $-326 \pm 90$ \\ 
\hline
K & 0.666 & -0.058 & $3.7 \pm 0.1$ & $75.8 \pm 1.2$ & $60.3 \pm 4.5$ & $236 \pm 20$ & $3.3 \pm 0.2$ & $44.1 \pm 0.8$ & $34.2 \pm 2.9$ & $121 \pm 14$ \\ 
L & 0.674 & -0.065 & $1.9 \pm 0.4$ & $45.0 \pm 1.8$ & $68.1 \pm 9.7$ & $139 \pm 35$ & --- & --- & --- & --- \\
M & 0.683 & -0.034 & $10.6 \pm 0.5$ & $67.3 \pm 0.2$ & $19.0 \pm 0.9$ & $215 \pm 14$ & $5.0 \pm 0.4$ & $67.3 \pm 0.6$ & $65.9 \pm 3.5$ & $354 \pm 36$ \\ 
\hline
N & 0.640 & -0.092 & $11.7 \pm 0.2$ & $82.4 \pm 0.2$ & $44.9 \pm 0.9$ & $559 \pm 15$ & $4.2 \pm 0.2$ & $39.5 \pm 0.6$ & $40.0 \pm 2.4$ & $180 \pm 14$
\enddata
\end{deluxetable*}

Figures~\ref{fig:sgrbregion} and \ref{fig:ciico} provide overviews of
the spatial and spectral characteristics of \cii\ emission toward the
\sgrb\ region.  Spatial and velocity continuity in \cii\ indicate that
the \sgrbo\ and the \gsix\ \hii\ region are part of a larger structure
that extends to positive Galactic longitude (\pell) to behind and
beyond \sgrbt.  As seen in \cii, the distinction between these
different regions appears to be rooted in the resolution of historical
single-dish continuum surveys \citep[e.g.]{mezger76}, rather than the
separation of physically disconnected structures.

Figure~\ref{fig:sgrbregion} is an image of \ciiw\ line intensity in
K\,\kms\ integrated over $14$ to $120$\kms.  For a characteristic
distance of 8\,kpc to the main \sgrb\ cloud, the \sgrb\ \cii\ region
stretches approximately 34\,pc along Galactic longitude $\ell$, and
15\,pc in latitude $b$.  The bright emission contained within the
175\,K\kms\ (lowest) contour around the main body, encompasses
465\,pc$^2$.

All of the main \cii\ emission peaks are associated with bright radio
continuum emission, although the converse is not always true.  The
brightest \cii\ regions, marked by circles A and B, are within the
radio continuum sources named the ``ionized rim'' and ``ionized bar,''
respectively, by \citet{mehringer92}.  The peak at $\ell = 0.60\degr$
(region D) is the \gsix\ \hii\ region, with its associated bright arc
visible below (region E is in the arc).  Crosses in the figure
indicate the locations of the luminous compact star forming cores
\sgrbt(N), \sgrbt(M) (region J), and \sgrbt(S).  As we discuss in
Sec.~\ref{ssec:missing}, none of the cores emit detectable \cii\
emission.

Figure~\ref{fig:ciico} compares the \cii, \thco $J = 2-1$, and \ceio\
$J = 2-1$ \citep{apexco} spectra averaged over the main body of the
\sgrb\ cloud indicated by the 175\,K\kms\ contour at the center of
Fig.~\ref{fig:sgrbregion}. The \cii\ line peaks near $50$\kms, with
emission from about $-30$ to $+130$\kms.  There appear to be two main
components, and a two-component Gaussian fit gives good estimates of
these, separating emission into a main component centered at
$46.0 \pm 1.0$\kms, with $5.3 \pm 0.03$\,K peak temperature and
$57.3 \pm 2.8$\kms\ width; and another at $85.5 \pm 0.27$\kms, with
$1.9 \pm 0.06$\,K peak and $71.5 \pm 0.80$\kms\ width.  Absorption
from the Galactic plane blocks \cii\ emission around 0\kms, masking a
likely wing to lower velocity.  The $\sim$50\kms\ component is
associated molecular and ionized gas \citep{apexco, mehringer92}, and
contains 69\% of the total \cii\ flux in this decomposition.  The
subsidiary peak at near $90$\kms\ is considerably more prominent in
extended CO emission than in \cii.

Overall agreement between \cii, \thco, and \ceio\ velocity structures
near $+50$\kms\ indicates that \cii\ and molecular emission are
associated to a large degree.  All of these lines, as well as the
H\,110$\alpha$ radio recombination line \citep{mehringer92}, peak
close to $+50$\kms, indicating that the \cii\ emission is associated
with the main column of molecular gas and its ionized surface.
Comparison of the CO lines suggests that the line shapes are dominated
by large-scale motions, and that these lines are not very optically
thick on average.

The $+90$\kms\ \cii\ emission is present across much of the image.  It
is likely associated with extended, moderately excited, molecular
material previously identified by \citet{vogel87} toward \sgrbt.
Excitation and velocity suggest that the emission is generally
associated with the Galactic center rather than the Galactic disk.
APEX CO and isotopologue $J = 2-1$ data cubes show an emission
component around $+90$\kms\ that stretches across the entire CMZ.

Figures~\ref{fig:chanmaps} and \ref{fig:pv} show that the velocity
field across the bright body of \sgrb\ has overall coherence in a
smooth increase in velocity with increasing Galactic longitude, along
with a great deal of fine structure.  Linewidths of 30 to above
50\kms\ are common across the region in individual 0.55\,pc beams,
indicating that the \cii-emitting material shares in the broad
linewidths typical for the Galactic center's dynamic molecular clouds.

Figure~\ref{fig:chanmaps} is a set of velocity channel maps in 20\kms\
bins, with the 70\um\ continuum image and contours for comparison.
The $+50$\kms\ \cii\ component is brightest toward \sgrbo's radio
continuum and 70\um\ peaks \citep{mehringer92, lang10, molinari16},
with additional lower-level emission around this velocity across the
entire \cii-bright region.  The channel maps also show a very wide
linewidth in the emission arc wrapping around \gsix\ to negative
latitude ($-b$).  The arc is visible at a variety of wavelengths, and
may be material swept up and compressed by winds from \gsix's central
luminosity sources.  In addition to extended $+90$\kms\ emission,
$+90$\kms\ \cii\ is also bright along a North-South band of wide
linewidth emission near the \gsix\ \hii\ region at $\ell = 0.60\degr$.

Overall velocity trends are clear in Figure~\ref{fig:pv}, a
position-velocity diagram along a line cut through a bright \cii\ peak
(region A), and through and beyond \sgrbt(M) (region J).  The mean
velocity of the main component of \cii\ emission across the entire
\sgrb\ body follows a smooth trajectory in the sense of Galactic
rotation, velocity increasing with $\ell$.  Lineshapes broaden and
become more complex at \gsix\ and beyond.  Bright 158\um\ continuum
associated with a North-South dust lane and perhaps the \sgrbt\ cores
at $\ell \approx 0.66\degr$ causes strong absorption from material in
the Galactic plane at $+7$\kms.  A faint streamer stretching in
velocity from $+50$ to over 110\kms\ from $\ell \approx 0.53\degr$ to
$0.55\degr$ in Fig.~\ref{fig:pv} is from line wings with velocities
above 100\kms\ visible in spectra between \sgrbo\ and \gsix.

Overall velocity trends are clear, but the velocity field is complex
in detail.  Figure~\ref{fig:spectralstack} explores the complexity by
comparing lineshapes from 15\asec\ diameter samples across \sgrb, with
locations keyed to the letters in Fig.~\ref{fig:sgrbregion}.  Our
convention is to set baselines to zero brightness, so line absorption
against continuum appears as negative-going structure.  Double
Gaussian component fits to the lineshapes provide good representative
summaries of the main emission; fit parameters are in
Table~\ref{tab:fits}.

\sgrb's most intense emission, in regions marked A and B, is
associated with the bright \sgrbo\ ``rim'' and ``arc'' radio continuum
regions near $\ell = 0.50\degr$.  Their \cii\ velocities of 44 and
47\kms\ agree well with H\,110$\alpha$ recombination line velocities
\citep{mehringer92}. Position A's linewidths are 49 and 20\kms, but B
is better fit by a single 33\kms\ FWHM component.  Both peaks A and B
have correspondence in 70\um\ continuum structures, but none in
160\um.

Continuing to positive longitude, a spectrum sampling the body of the
cloud at position C has a 32\kms\ FWHM component at 53\kms.  The main
\cii\ line component from the \gsix\ \hii\ region at position D has a
center velocity of 50\kms\ and width 33\kms\ FWHM, with a distinct
second component at the substantially higher velocity of 84\kms, with
57\kms\ FWHM. As is typical for all but the \sgrbt\ cores, \cii\ and
\ceio\ lineshapes share similarities, indicating that the spectrum
identifies two separate structures rather than a single broad line
with foreground absorption at an intermediate velocity.  \gsix\ has
what appears to be a partial shell, appearing as an arc toward
negative latitude; the spectrum at position E is representative along
the arc, with a 26\kms\ FWHM peak near 80\kms\ matching the component
in \gsix\ itself, and a very broad (93\kms\ FWHM) wing to lower
velocity, centered at 62\kms.

Regions F through I sample emission along the dust lane that crosses
most of \sgrb's body.  The \sgrbt(M), (N), and (S) cores are all close
to the \pell\ (eastern) side of this high column density dust
structure that matches the size and shape of the \cii-dark notch and
\cii-dim stripe visible in Fig.~\ref{fig:sgrbregion}.  As we discuss
in Sec.~\ref{ssec:comparisons}, this approximately North-South strip
is a clear feature in infrared continuum images, becoming increasingly
darker and longer from 70\um\ to 8\um\ wavelengths.  At 160\um, which
is representative of continuum underlying the 158\um\ \cii\ line, the
lane is bright at its northern end, encompassing regions G-J and N,
with peaks at the \sgrbt\ cores, fading in brightness to the south
before dropping to the typical cloud brightness level just beyond
region F.  All \cii\ spectra along this \cii-dim and IR-dark lane have
similar lineshapes dominated by a 65 to 70\kms\ FWHM component
centered near 67\kms.  All but the southernmost position F have
sufficient 158\um\ continuum to cause absorption from about 0 to
20\kms\ in the Galactic plane.  \cii\ emission's lineshape similarity
and run of peak line brightness with dust column density suggest that
the spectra originate in a generally cool dust lane that may be
optically thick to \cii\ from emission behind it. It appears that the
material producing the notch and dark lane are between us and the
continuous body of \sgrb\ otherwise highlighted in \cii.

The most dramatic spectrum is at position J, coincident with the
maximum absorption toward \sgrbt(M) and the peak of 1.3\,mm continuum
measured by ALMA \citep{sanchezmonge17}.  Ignoring Galactic plane
absorption from $-60$ to $+$20\kms\ visible against \sgrbt(M)'s
intense continuum, emission in this spectrum is best represented as a
66\kms\ FWHM line centered at 68\kms, with a 37\kms\ FWHM absorption
to approximately zero brightness centered at 70\kms.  We discuss this
feature and its implications in detail in Secs.~\ref{ssec:physstruct}
and \ref{ssec:missing}.  While \cii\ absorption from the Galactic
plane is strongest against \sgrbt(M), and is slightly enhanced by
increased 158\um\ continuum toward \sgrbt(N) and (S), it extends along
the entire 160\um-bright (and \cii-faint) North-South strip visible in
Fig.~\ref{fig:sgrbregion}.  With the exception of the 70\kms\
absorption notch, the emission lineshape toward J is very similar to
surrounding positions including I, which is in the darkest part of the
dark lane at shorter wavelengths, and brightest at 160\um.

Velocity components near 70 and 50\kms\ are present throughout the
regions near \sgrbt(M) and to larger longitudes (see also
Fig.~\ref{fig:chanmaps}).  The bright spot at position M, near
\sgrbt(M), has both 19 and 66\kms\ FWHM lines centered at
67\kms. Further to $+\ell$, region K has components centered at both
44 and 76\kms, and region L peaks at 45\kms. Sampling to $-b$, the
bright region K has two velocity peaks, one at 76\kms\ with 60\kms\
FWHM, and the other at 44\kms\ with 34\kms\ FWHM. The run of nearby
spectra shows that these are separate emission peaks, and not the
residuals from absorption against a broader line at intermediate
velocity.  As with other \cii\ peaks, this region is close to a local
peak of 20\,cm \citep{lang10} and 70\um\ continua, but has no 160\um\
\citep{molinari16} enhancement.

\begin{figure*}[!ht]
\centering
\includegraphics[width=0.65\textwidth]{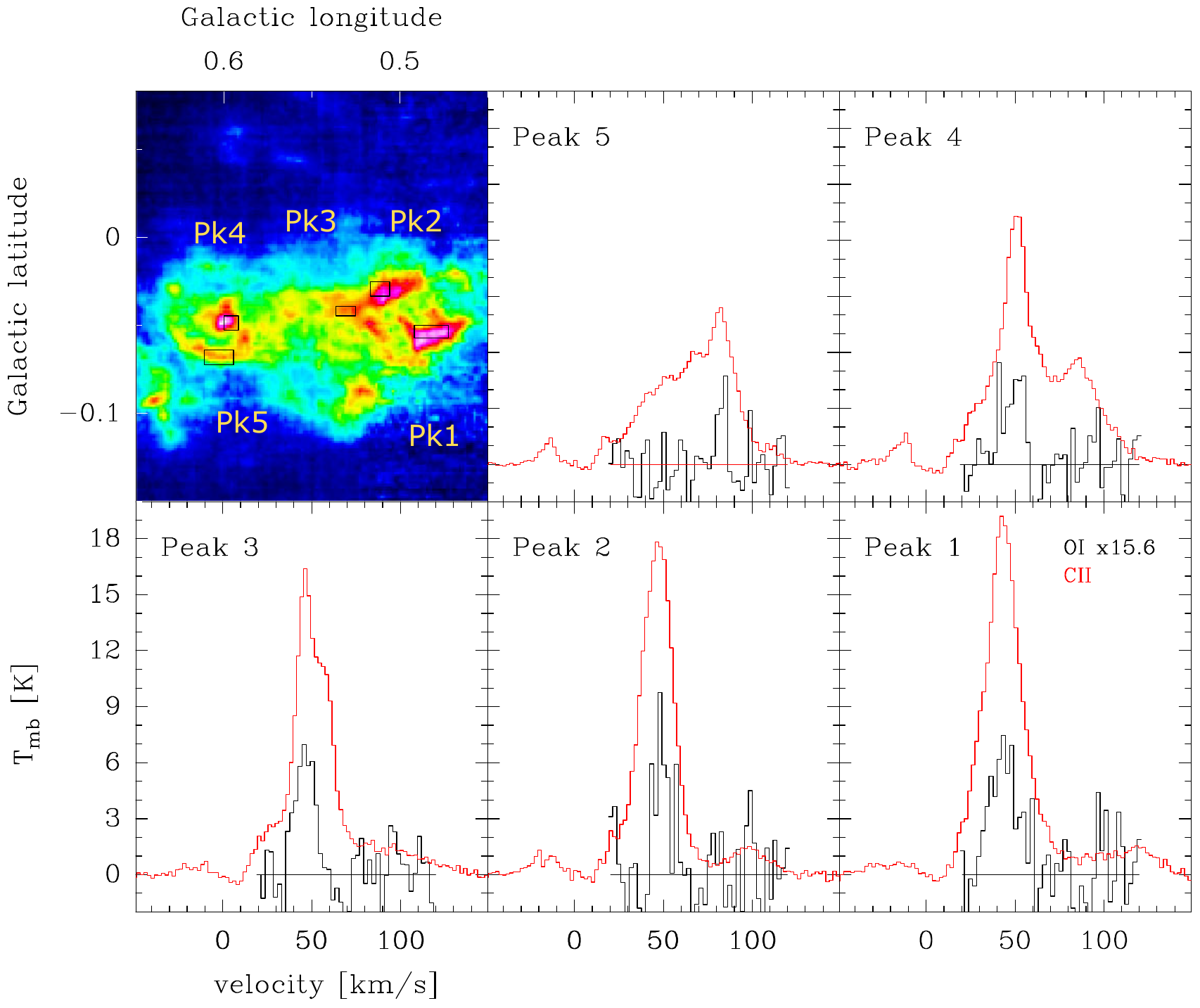}
\caption{
Integrated \cii\ intensity (top left) and \cii\ and \oi\ spectra
averaged over black rectangles in bright regions, labeled in
increasing order with longitude (right to left in image).  The \oi\
brightness temperature scale has been multiplied by
(157.74\um//63.18\um)$^3 = 15.6$ to allow direct comparison of
intensities.  The intensity ratio is $I(\mbox{\oi})/I(\mbox{\cii})
\approx 0.3$ (see Table~\ref{tab:oiIntens}), in agreement with typical
values for unresolved spectra across the region \citep{goi04}.  This
implies similar underlying physical conditions in the
\oi-emitting material across the source, on average.     
\label{fig:ciioispect}}
\end{figure*}

\subsection{Additional \cii\ components}\label{sec:compact}

The \cii\ data cube reveals a number of isolated clouds at velocities
other than those that are typical for \sgrb's body.  The most prominent are:

A distinct 40\asec\ diameter source with a 9.2\,K peak, 6.5\kms\ FWHM
line toward \ra{17}{47}{18.6}, \dec{-28}{40}{22}
($\ell = 0.417\degr, b = -0.178\degr$).  At an LSR velocity of
$-58.7$\kms, this source is likely associated with the so-called
3\,kpc expanding arm.  The source appears as a bright spot at the
southern end of a small \thco\ $J = 2-1$ filamentary cloud with the
same linewidth and velocity.  A lack of a radio continuum counterpart
in the \citet{lang10} 20\,cm continuum data implies the absence of an
\hii\ region, in line with the general lack of compact \hii\ regions
in the 3\,kpc arm.  Other similarly narrow line components in the
\cii\ data cube fall close in velocity to the Galactic plane
absorption, but are most likely remnants of the wings of broader
self-absorbed lines.

A small elliptical cloud with size 37\asec\ by 77\asec\ at PA
$330\degr$ and peak $T_{MB} = 1.6$\,K, velocity 124\kms, and width
36\kms\ is centered at \ra{17}{46}{55.5}, \dec{-28}{18}{56} ($\ell =
0.679\degr$, $b = 0.078\degr$).  \thco\ and \ceio\ $J = 2-1$ in the area show
mainly emission peaking at about $+40$\kms.  The bulk of \cii\ here is
from material with very low CO emissivity: while \thco\ at velocities
near $128$\kms\ is present in patches throughout the region, it
matches only the edge of the higher-velocity \cii\ wing, and \ceio\ is
not clearly present. 

\subsection{\oi\ spectroscopy\label{ssec:oi} }

Figure~\ref{fig:ciioispect} shows the \grt\ velocity-resolved \oisw\
spectra from the brightest \cii\ regions.  \oi\ emission is mostly at
the $\sim$50\kms\ velocity of the \sgrb\ cloud, and shares peak
velocities with \cii\ even when multiple components are visible in the
\cii\ line.  Coincidence between \cii\ and \oi\ indicates that \cii\
is associated with neutral gas, and that emission from photodissociation
regions (PDRs) may dominate in the brighter, narrower-line \cii\
component.  The peaked \oi\ lineshapes suggest that the lines
are not self-absorbed, so \oi\ intensities are accurate measures of
\oi's contribution to gas cooling.  

Comparing line intensities gives
$I(\mbox{\oi})/I(\mbox{\cii}) \approx 0.3$ (Table~\ref{tab:oiIntens}).
Ratios from Table~\ref{tab:oiIntens} are equivalent to the typical
value of 0.3 from unresolved {\em ISO}-LWS spectra in $\sim$80\asec\
beams \citep{goi04} across the \sgrb\ cloud.  In our data, only the
brightest \cii\ regions show the weaker \oi\ clearly. Our mapping
strategy was driven by the signal to noise ratio for the \cii\ line;
residual spectral baseline structure in the \oi\ spectra prevented
extraction of lower-level \oi\ emission averaged over larger
regions. The similar intensity ratio with the {\em ISO} data suggests
that \oi\ scales with \cii\ in fainter regions as well, however,
pointing to a component of \cii\ with similar physical conditions and
associated with neutral gas sufficiently dense
($n \gtrsim {\rm few} \times 10^3$\percmcu) to excite \oi\ across the
entire bright region.

\begin{deluxetable}{rrrrrrrr}[!ht]
\tabletypesize{\scriptsize}
\tablecaption{\cii\ and \oi\ intensities and intensity ratios for the
regions indicated in black rectangles in Fig.~\ref{fig:ciioispect}.
Center positions and region sizes are in Galactic coordinates.
Intensities are in units of $10 ^{-3} \;
[\mathrm{erg~s^{-1}~cm^{-1}~sr^{-1}}]$.
\label{tab:oiIntens}}
\tablewidth{0pt}
\tablecolumns{8}
\tablehead{
\colhead{Peak} &
\colhead{$\ell$} &
\colhead{$b$} &
\colhead{Size} &
\colhead{Vel.} &
\colhead{$I$(\cii)} &
\colhead{$I$(\oi)} &
\colhead{$I$(\oi)/} \vspace{-3mm} \\
\colhead{} &
\colhead{[deg]} &
\colhead{[deg]} &
\colhead{[arcsec]} &
\colhead{[km/s]} &
\colhead{ } &
\colhead{ } &
\colhead{$I$(\cii)}
}
\startdata
1 & 0.482 & $-0.053$ & $ 70 \times 25 $ & 25--60 & 2.99 & 0.87 & 0.29 \\ 
2 & 0.511 & $-0.029$ & $ 40 \times 30 $ & 40--60 & 1.96 & 0.57 & 0.29 \\ 
3 & 0.531 & $-0.042$ & $ 40 \times 20 $ & 35--60 & 1.93 & 0.64 & 0.33 \\ 
4 & 0.596 & $-0.049$ & $ 30 \times 30 $ & 40--60 & 1.41 & 0.42 & 0.30 \\ 
5 & 0.603 & $-0.068$ & $ 60 \times 30 $ & 75--90 & 0.75 & 0.15 & 0.21
\enddata
\end{deluxetable} 

\begin{figure*}[!ht]
\centering
\includegraphics[height=8.4in]{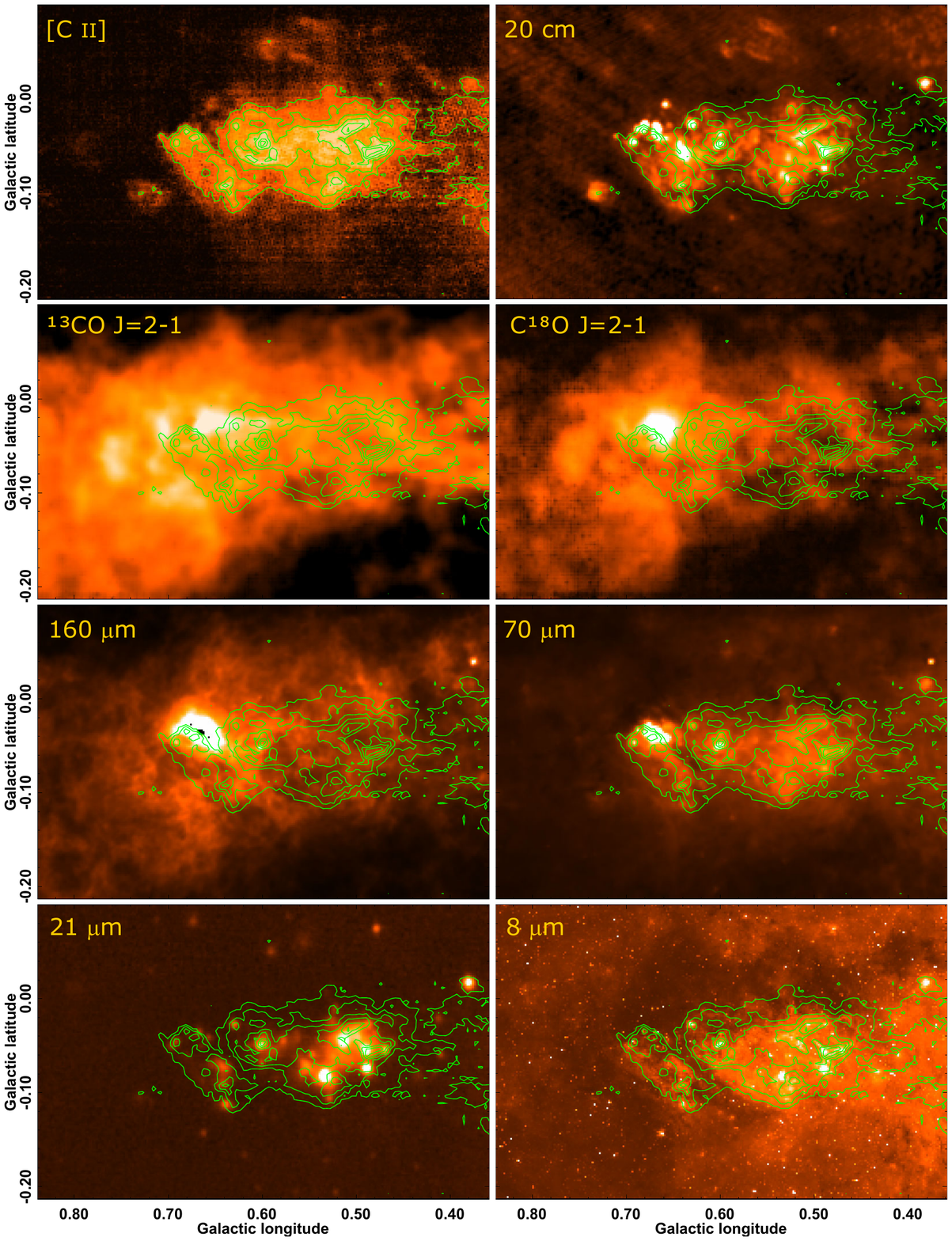}
\caption{
Integrated intensity images of various tracers toward \sgrb. \cii\
contours from Fig.~\ref{fig:sgrbregion} are on images of 20\,cm radio
continuum \citep{lang10}, APEX CO isotopologues $J=2-1$ integrated
over $\pm 120$\kms\ \citep{apexco}, {\em Herschel/PACS} 160\um\ and
70\um\ continuum \citep{molinari16}, and 21\um\ {\it MSX} and 8\um\
{\em Spitzer} \citep{price01, stolovy06} continuum, all sampled to the
\cii\ beamsize. Color stretches are linear, trimmed to emphasize
structure.  Galactic latitude and longitude axes are in degrees.
\cii, 70\um, and 20\,cm intensities correspond closely, with the
exception of the \sgrbt\ cores at $\ell \approx 0.68\degr$, while the
background cloud traced in CO and 160\um\ is much larger than the
\cii\ region, and extends considerably further to
$+\ell$. Mid-infrared images generally emphasize compact sources.
\label{fig:Integ}}
\end{figure*} 

\section{Discussion} \label{sec:discus}

With 12\% of the \cii\ emission from only 6\% of the area in our large
scale map of the entire CMZ \citep{wholecmz}, the \sgrbo\ region is a
notable contributor to the entire \cii\ line flux emitted by the
Galactic center region.  The \sgrbo\ region is second only to the
\sgra-Arches region as the brightest source of \cii\ across the entire
CMZ, with the difference due to \sgrb's smaller extent rather than
lower surface brightness.  Our data cube with sub-parsec spatial and
1\kms\ spectral resolution over an area of 3000\,pc$^2$ provides
context to understand the \cii\ excitation within the Galactic center,
as well as serving as a case study for interpreting \cii\ emission
from other galactic nuclei.

\subsection{Following the UV from young stars \label{ssec:comparisons}}
Ultraviolet radiation from young stars produces bright \cii\ and heats
dust as the surfaces of nuclear molecular clouds intercept and convert
essentially all UV to longer wavelengths.  \cii\ is therefore expected
to be a good spectroscopic tracer of star formation in obscured
regions throughout the universe (e.g., \citealt{stacey91, stacey10,
  delooze14, pineda14, herreracamus15}).  SOFIA-\grt's resolution and
the Galactic center's proximity allow us to examine the relationship
between \cii\ and far-IR luminosities within a galactic nucleus in
detail.  Imaging of \sgrb\ is at a scale between small isolated clouds
conducive to modeling and scales where relationships between the
far-UV from star formation and indirect tracers on square kiloparsec
scales hold (e.g., \citealt{calzetti10, kennicutt12}).

Figure~\ref{fig:Integ} shows comparisons of a number of potential
indirect tracers of far-UV.  The panels all show \cii\ (contours) with
integrated intensity images of 20\,cm radio continuum \citep{lang10},
APEX CO isotopologues $J=2-1$ integrated over $\pm 120$\kms\
\citep{apexco}, {\em Herschel}/PACS 160\um\ and 70\um\ continuum
\citep{molinari16}, and {\it MSX} 21\um\ and {\it Spitzer} 8\um\
\citep{price01, stolovy06} continuum, all sampled to the \cii\
beamsize.  Visually, the 20\,cm and 70\um\ distributions appear most
similar to \cii's morphology. \thco, \ceio, and 160\um\ are more
extended, especially to $+\ell$.  The mid-IR images have yet different
distributions, more concentrated toward the \cii\ peaks.

\begin{figure}[!ht]
\centering
\includegraphics[width=0.3\textwidth]{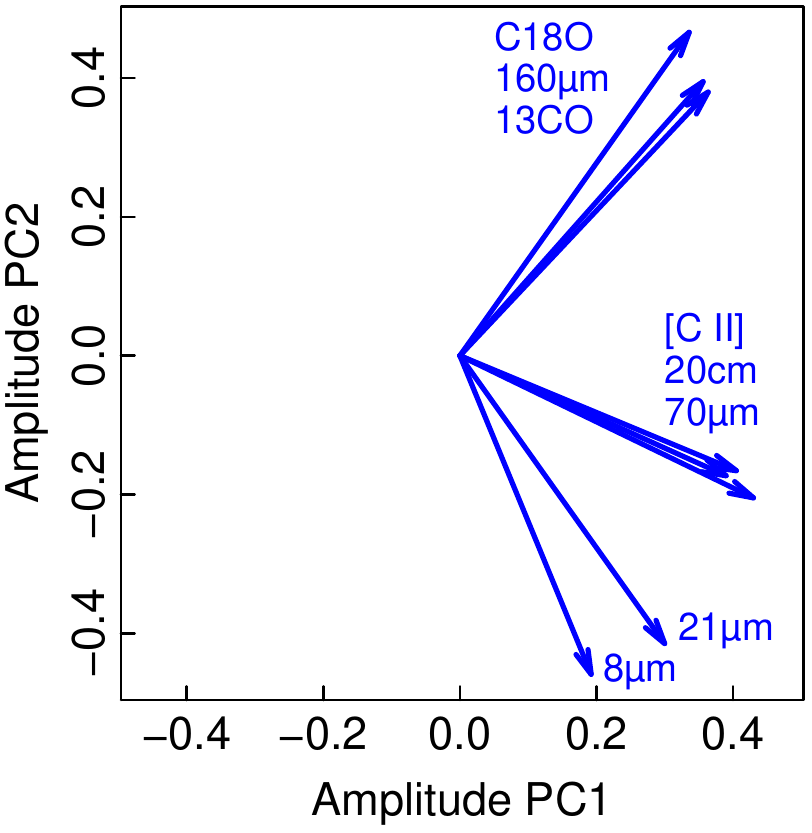}
\caption{
Amplitudes of the first two Principal Component vectors, PC1 and PC2,
from a decomposition of continuum and integrated intensity images for 8
different tracers.  Three groupings show spatially related distributions
from the large-scale cloud (\ceio, 160\um, and \thco), a tight cluster
tracing UV radiation from the \sgrb\ body (\cii, 20\,cm, and 70\um),
and the mid-IR tracing UV and compact sources (21\um, 8\um). The
decomposition excludes a small area immediately covering the \sgrbt\
cores, which has little effect on the result; see text.  
\label{fig:biplot}}
\end{figure}

To put the spatial comparisons on a quantitative basis,
Figure~\ref{fig:biplot} shows the amplitudes of the first two
components in a principal components analysis (PCA) decomposition of
this set of images.  PCA is a standard tool of multivariate analysis,
long used for multi-spectral imaging, either in different wavebands or
across spectra (\citealt{frieden91, ungerects97, heyer97} among
others, contain tutorials and examples).  PCA constructs an orthogonal
set of basis vectors from the data themselves, here the concatenated
and normalized columns of each image.  Any image can be reconstructed
from a linear combination of the new basis vectors.  After arranging
the principal components (PCs) in decreasing order of variation, the
first principal component (PC1) gives the most common vector (or
image); orthogonality ensures that the second PC (PC2) shows the
largest difference from PC1 and is independent from it; PC3 shows the
largest independent differences from both PC1 and PC2; and so on.
Higher principal moments typically contain little information other
than noise.  Vectors with similar principal component amplitudes
highlight similar structure within the vectors, or equivalently, in
the images.

Figure~\ref{fig:biplot} is a plot of the amplitudes of the PC1 and PC2
vectors from the 8 images.  The first two principal components contain
80\% of the total variation among the images, so considering these two
PCs reveals the major commonalities and differences between the
images.  All the vectors in Fig.~\ref{fig:biplot} have a positive PC1,
indicating a similar overall structure, but their PC2s, showing
deviations from the common structure, form three distinct groups. The
uppermost group contains \ceio\ $J = 2-1$, 160\um, and \thco\ $J =
2-1$.  The middle group shows a very tight cluster of \cii, VLA
20\,cm, and 70\um\ emission.  Infrared 21\um\ and 8\um\ images form a
looser third group.  This confirms the visual impression from
Fig.~\ref{fig:Integ}: while all images trace a roughly similar
structure, the top group shows more extended emission than the others
(Fig.~\ref{fig:Integ}), tracing large background cloud emission, while
the bottom group has high contrast between extended emission and
bright compact sources.  With a large PC1 compared with a smaller and
negative PC2, the middle group of \cii, 20\,cm, and 70\um\ is more
typical of the global emission, although with somewhat more of the
compact than the extended emission.  For full sensitivity to the
extended emission, this decomposition blanked a region $\Delta \ell =
130$\asec\ by $\Delta b = 100$\asec\ centered at \ra{17}{47}{21},
\dec{-28}{28}{03} ($\ell, b = 0.669, -0.038$) in all data sets.  This
region covers the \sgrbt\ cores and their immediate surroundings,
while retaining more than 95\% of the total flux in all of the images.
Without blanking, the middle set of vectors spreads to some extent,
while still retaining their grouping, but there is no effect on the
ordering, and little effect on the top and bottom groups.

\begin{figure*}[t]
\centering
\includegraphics[width=\textwidth]{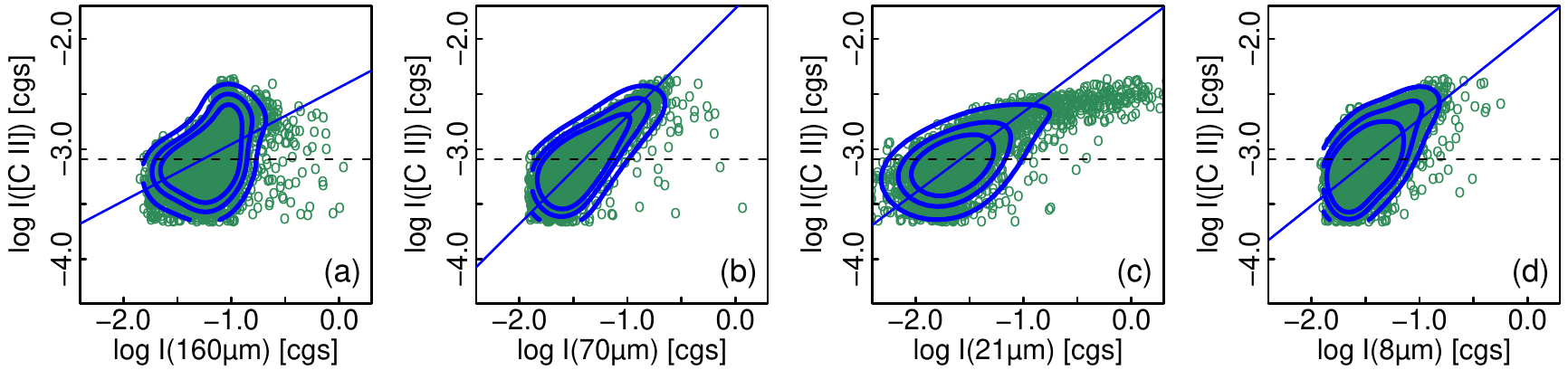}
\caption{
Intensity-intensity correlations of integrated \cii\ intensity vs.\
160\um, 70\um, 21\um, and 8\um\ continuum intensities across the
\sgrb\ region. Intensity units are cgs,
$[\mathrm{erg~s^{-1}~cm^{-1}~sr^{-1}}]$. The \cii\ emission has been
integrated from 14 to 120\kms.  Each point (in green) is averaged over
an independent 21\,arcsecond square region, with contours showing the
density of points (in blue) at 0.3, 0.6, and 0.9 of the peak density
of points.  The straight lines are linear fits to the density of data
points, weighted by the density to discriminate against outlier
points.  Table~\ref{tab:slopeInt} contains slopes and
intercepts. Scales are all 2.5 by 2.5 dex, and the horizontal dashed line
marks the 175\,K\,\kms\ \cii\ intensity cut for the \sgrbo\ body  
indicated by the lowest contour line in Fig.~\ref{fig:sgrbregion}.
\label{fig:ciiDust}}
\end{figure*}

Another quantitative view of the spatial relationships between
different wavebands is the spatial correlation of \cii\ with dust
continuum in bands frequently used to trace star formation.
Figure~\ref{fig:ciiDust} shows such correlations.  Each point in the
plots represents an average over an independent 21\asec\ region with
signal detected above $3\sigma$ in both tracers. Contour lines show
the density of points, with the straight line giving a linear fit to
the density of data points, weighted by the density of points to
discriminate against outliers.  Table~\ref{tab:slopeInt} contains fit
results.  In all panels, the sprays of points to higher infrared
intensities lacking strong correlation with \cii\ intensities are from
compact and IR-luminous regions with unrelated \cii\ emission.  The
correlation is tightest for \cii-70\um, panel (b), with the same
relationship within the body of \sgrb\ and in the fainter regions
outside that.

\begin{deluxetable}{rrrr}[!ht]
\vspace{5mm}
\tabletypesize{\scriptsize}
\tablecaption{Power laws and intercepts from linear fits to intensity
  correlation data of Fig.~\ref{fig:ciiDust}, of form
  $\log_{10}(\{I(\mbox{\cii})\}) = a \log_{10}\{I({\rm IR})\} + b$.
  Uncertainties are 68\% confidence intervals from bootstrap 
  iterations of weighted fits. The last column is the Pearson 
  correlation coefficient $r$. \label{tab:slopeInt}}
\tablewidth{0pt} \tablecolumns{3}
\tablehead{ 
\colhead{IR band} & 
\colhead{$a$} & 
\colhead{$b$} &
\colhead{r}
}
\startdata
160\um & $ 0.52 \pm 0.08 $ & $ -2.44 \pm 0.10 $ & 0.44 \\ 
70\um & $ 0.98 \pm 0.05 $ & $ -1.72 \pm 0.06 $ & 0.89 \\ 
21\um & $ 0.73 \pm 0.08 $ & $ -1.93 \pm 0.13 $ & 0.72 \\ 
8\um & $ 0.78 \pm 0.07 $ & $ -1.94 \pm 0.10 $ & 0.66
\enddata
\end{deluxetable}

Correlation is poor for 160\um-\cii, panel (a), as the bright
continuum luminosity is concentrated around the \sgrbt\ cores and the
northern end of the dark lane, while \cii\ is bright across the \sgrb\
body.  The brightest \cii\ regions do not have corresponding peaks at
160\um, and vice versa.  At lower intensities, 160\um\ traces the
column density of the larger background cloud that is also apparent in
CO isotopologues.  Some hint of two clusters of points in the
correlation plot may be attributable to separation between bright and
background emission.

The best-calibrated mid-IR tracer for obscured star formation over
large scales, {\em Spitzer} 24\um\ \citep{calzetti10}, had saturated
regions that prevented comparisons over the entire field, so we show
{\it MSX} 21\um\ data as an alternative in panel (c). The distribution
is tadpole-shaped, showing a general relationship between much of the
emission at the two wavelengths, with a flatter power law indicating
low correlation between bright 21\um\ emission from the compact
sources and \cii.  The 8\um-\cii\ correlation is in panel (d).  The
broader distributions about the mean compared with 70\um\ reflect the
differences in spatial distributions.  A much tighter correlation
between 8\um\ and \cii\ intensities in the Orion region
\citep{pabst17} is attributed to UV excitation of both 8\um\ bands of
polycyclic aromatic hydrocarbons (PAHs) and \cii\ in PDRs.  Unlike
those sources, which are largely isolated and extended diffuse clouds,
the \sgrb\ field contains a mix of extended and compact sources, and
is affected by absorption and emission from the $\sim$8\,kpc of
Galactic plane along the line of sight.

\begin{figure}[!ht]
\centering
\includegraphics[width=0.25\textwidth]{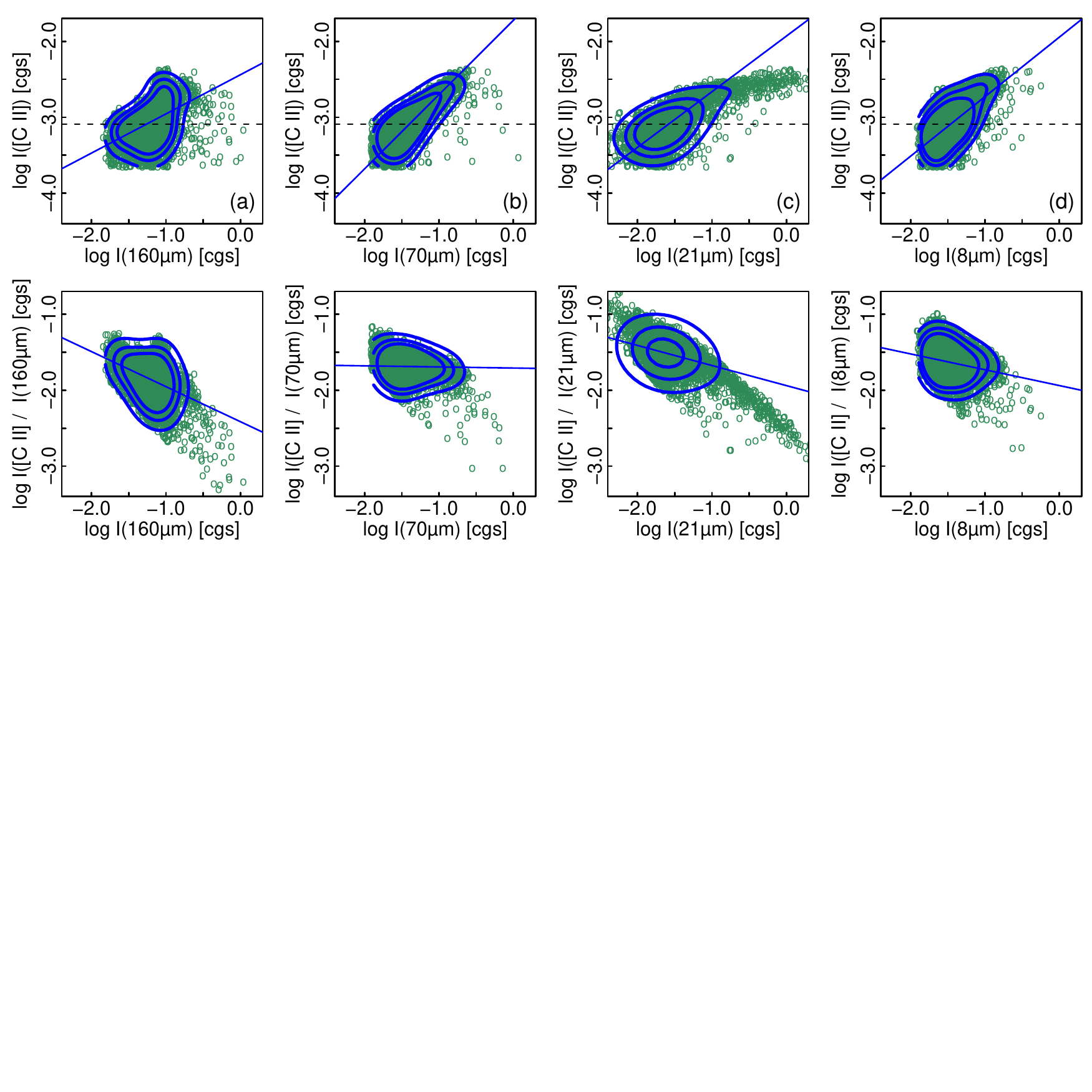}
\caption{ \cii\ emission efficiency, measured as the ratio of \cii\ to
  70\um\ intensities, vs.\ 70\um\ intensity. Data and fitting method
  are the same as in Fig.~\ref{fig:ciiDust}, yielding a slope of
  $ -0.01 \pm 0.07 $, or negligible dependence on 70\um\ intensity.
  Since 70\um\ is a proxy for FIR intensity within the Galactic center
  PDRs (Sec.~\ref{sec:physcond}), and a scaling factor would affect
  both axes equally, this plot also shows that the heating efficiency
  is independent of FIR intensity.
\label{fig:ciiHeating}}
\end{figure}

Figure~\ref{fig:ciiHeating} provides a view of the relationship
between \cii\ and 70\um\ intensities across \sgrb.  Using the same
data and fitting method as for Fig.~\ref{fig:ciiDust}, we find a slope
of $ -0.01 \pm 0.07 $, indicating that radiation and physical
conditions toward \sgrb\ do not appreciably suppress the conversion of
UV to \cii.  \cii\ and 70\um\ trace UV deposition at the cloud surface
equally well across \sgrb's body, with the exception of the brightest
emission associated with the \sgrbt\ cores.  70\um\ emission is near
the peak of a graybody dust emission curve (Sec.~\ref{sec:physcond})
and is tightly correlated with \cii, making it an accurate proxy for
the total FIR intensity from the \cii-emitting regions in \sgrb.  A
scaling factor between 70\um\ and FIR intensities would change axis
offsets but not the relationship's slope in Fig.~\ref{fig:ciiHeating},
indicating that the heating efficiency, measured as the ratio of \cii\
to FIR intensities, vs. FIR intensity, is also nearly constant.

All of the comparisons above show a close connection between \cii\ and
70\um\ continuum, reinforcing the interpretation that they both arise
from UV depositing energy at cloud edges.  The comparisons also show
that 20\,cm VLA imaging at $\sim$15\asec\ resolution toward the
Galactic center \citep{mehringer92, lang10} resolves out the Galactic
plane emission accounting for about half of the total flux toward the
center in single-dish observations \citep{law08}, while retaining good
sensitivity to the free-free emission from structures in the center.
Agreement between the far-IR and 20\,cm distributions also confirms
that the synchrotron contribution toward \sgrb\ at 20\,cm is small.
The other bands we considered do not pick out the far-UV as cleanly.
Emission at 160\um, \thco, and \ceio\ are products of temperature and
column density, and trace combinations of large columns of cooler
material in background clouds as well as warm surfaces. Emission from
8\um\ and 21\um\ in the Galactic center is weighted toward bright,
compact regions, with dust absorption obscuring the most luminous
cores, and line of sight material in the Galactic plane causing some
confusion for extended emission.  These results favoring \cii\ and
70\um\ as UV tracers likely hold in other galactic nuclei as well.  A
fortunate combination of circumstances for 20\,cm radio continuum
(i.e., the fraction of resolved large-scale emission, emission size
scale, and lack of significant non-thermal emission) may not apply
elsewhere in the Galactic center or in other galactic nuclei, however.

\subsection{Physical conditions in the \cii\ emission regions \label{sec:physcond}}

In this section we examine the emission from the bright body of \sgrb\
contained within the lowest contour in Fig.~\ref{fig:sgrbregion} to
derive overall properties.  A key question we approach is the fraction
of \cii\ emitted from \hii\ regions compared with photodissociation
regions (PDRs).  This is of intrinsic interest in understanding the
emission from the CMZ, and it is also a guide to interpreting \cii\
emission from external galaxies.

We set a lower limit on the column density and mass of singly ionized
carbon in the bright \cii\ region of \sgrb's body by assuming that the
ions are in collisional equilibrium with their surroundings ($n >
n_{crit}$ of a few thousand cm$^{-3}$ in molecular hydrogen gas, or an
order of magnitude lower in ionized hydrogen plasma;
\citealt{launay77, flower77, wiesenfeld14}); that the temperature is
well above the level energy (91\,K); and that the emission is
optically thin. From expression (A2) in \citet{crawford85} (see
also \citealt{goldsmith12}), the measured \cii\ intensity of the bright
region outlined in Fig.~\ref{fig:sgrbregion} is $2.00\times 10^{-3}$
erg\,cm$^{-2}$\,s$^{-1}$\,sr$^{-1}$, corresponding to a \cp\ column
density $N_\mathrm{C^+} = 1.16\times 10^{18}$\percmsq\ and a mass
$m_\mathrm{C^+} = 50.0$\msun.  Following \citet{genzel90} to allow
direct comparisons with other Galactic center results, we take a
[C]/[H] abundance ratio of $3\times 10^{-4}$ and assume all carbon is
in \cp, deriving a hydrogen column density $N_\mathrm{H} = 3.86\times
10^{21}$\percmsq\ and a mass $m_\mathrm{H} = 1.39 \times 10^4$\msun.
These values are comparable with those \citet{genzel90} found for the
Arches region of the Galactic center.  A lower gas-phase carbon
abundance, such as that found along diffuse lines of sight in the
Galactic disk \citep{sofia04}, or with carbon in forms other than \cp,
would proportionally increase these estimates.

We use the \citet{lang10} 20\,cm VLA and \citet{law08} 21\,cm Green
Bank Telescope radio continuum data over the same region as the bright
\cii\ emission within the lowest contour of Fig.~\ref{fig:sgrbregion}
to constrain the \cii\ contribution from \hii\ regions.  We estimate
that about one half of the 20\,cm flux averaged over the entire region
is from the foreground Galactic plane by averaging across longitude
over the latitudes toward \sgrb.  Comparing VLA and GBT fluxes toward
\sgrb, we estimate that the VLA image contains 0.54 of the total
\sgrb\ flux, spatially resolving out the rest.  \citet{mehringer92}
report continuum and recombination line observations of compact
regions across \sgrbo\ and \gsix, which cover much of the \sgrb\
complex area.  They estimate that the 20\,cm emission is optically
thin with little contribution from synchrotron radiation in this
region, an estimate confirmed in multi-frequency single-dish
observations by \citet{law08}, and give typical electron temperatures
of $T_e \approx 4800$\,K for the compact regions.  With this electron
temperature and the \citet{lang10} flux density corrected for power
resolved out by the interferometer, the emission measure across
\sgrb's bright region is $EM = 1.72\times 10^{5}$\,pc\,cm$^{-6}$.  The
relationship between emission measure, column density $N$, and
particle density $n$ is $N \sim EM/n$, allowing an estimate of
$N$. Infrared fine-structure lines from the \sgrb\ \hii\ region
provide estimates of electron density of about 300\percmcu\
(\citealp{goi04, simpson18b}; this is also sufficient to thermalize
any \cii\ line emission from the \hii\ region).  Combining these
values, we find $N(\mathrm{H^+}) = 1.9 \times 10^{21}$\percmsq, for a
total ionized mass of $m_\mathrm{H^+} = 5.7 \times 10^3$\,\msun\ in
the region.  These densities imply that the ionized gas has a
characteristic line of sight depth of $s \approx N/n = 2$\,pc, which
seems plausible as a relatively thin layer over the region's surface.

Taking the values above, and given the uncertainties in the continuum
contribution from the Galactic plane, the H$^+$ filling factor, and
the atomic carbon abundance, we find a representative ratio of
$N(\mathrm{H^+})$/$N(\mathrm{H}) \approx 0.5$.  Most corrections push
this ratio to lower values: Although the \cii\ lines are still in the
``effectively optically thin'' regime found by \citet{goldsmith12},
there is some evidence for \cii\ optical depths of a few within the
CMZ \citep{wholecmz}, and substantial \cii\ optical depths would
decrease the ratio.  A lower gas-phase carbon abundance or increased
electron density would also decrease the ratio, as would synchrotron
contributions to the radio flux. Ionized hydrogen columns increase
sub-linearly with $T_e$.

Both PDRs and \hii\ regions emit \cii\ line radiation, but \oi\ is
excited in PDRs and destroyed in \hii\ regions.  PDR models combined
with observed \cii, \oi, and far-infrared continuum therefore provide
another constraint on the fraction of \cii\ emission from neutral
material.

We measure the \cii\ and \oi\ intensities directly, but we must
separate the PDR's far-infrared dust intensity from cooler and warmer
cloud components. Since 70\um\ is very similar in spatial distribution
to \cii\ (Sec.~\ref{ssec:comparisons}), we estimate $I$(FIR) by scaling
the intensity falling within the 70\um\ PACS filter passband to the
full 8--500\um\ FIR band from a
\begin{equation}
  I_\lambda(\lambda, \, T) \propto \left(1 -
    \exp\left[\left(\lambda/\lambda_o\right)^{-\beta}\right]\right)\,
B_{\lambda}(\lambda, \, T)
\label{eq:graybody}
\end{equation}
graybody.  \citet{goi04} used a two-component version of this
expression to fit far-IR continuum fluxes in the region.  They found
cool and warm dust components, with warm dust component temperatures
of 30--38\,K and $1 \leq \beta \leq 1.5$ in the bodies of the massive
clouds in their 80\asec\ beams.  

For the warmer dust associated with the PDRs, we prefer a temperature
of 45\,K $\beta = 1.5$, based on the tight similarity of the 70\um\
and \cii\ images and match to the 70\um/21\um\ flux ratio, including a
correction for PAH emission \citep{draine07}.

With this combination of temperature and $\beta$, a fraction 0.245 of
the total 8 to 500\um\ far-IR flux falls in the PACS 70\um\ filter
bandwidth, or $I({\rm FIR}) = I$(70\um)/0.245. The observed 70\um\
mean intensity of 0.11 erg\,s$^{-1}$\,cm$^{-2}$\,sr$^{-1}$ translates
to $I$(FIR) = 0.45 erg\,s$^{-1}$\,cm$^{-2}$\,sr$^{-1}$ from the PDR
regions.  This value is insensitive to modest changes in temperature
or $\beta$ because the 70\um\ band is on the Rayleigh-Jeans side close
to the graybody peak. For $\beta = 1.5$, $T = 35$\,K gives a fraction
of 0.262; while at $T = 45$\,K, $\beta$ of 1.25 to 1.75 gives 0.248 to
0.239.  Our derived intensity of $I$(\cii)/$I$(FIR)$= 0.46$\% is in
line with ratios of 0.1 to 1\% for starbursts \citep{stacey91} and
$0.48 \pm 0.12$\% for a large sample of disk galaxies \citep{smith17}.
We conclude that the 70\um\ flux can be scaled to provide a good
measure of the total FIR intensity for PDR modeling.

\begin{figure}[!ht]
\centering
\includegraphics[width=0.45\textwidth]{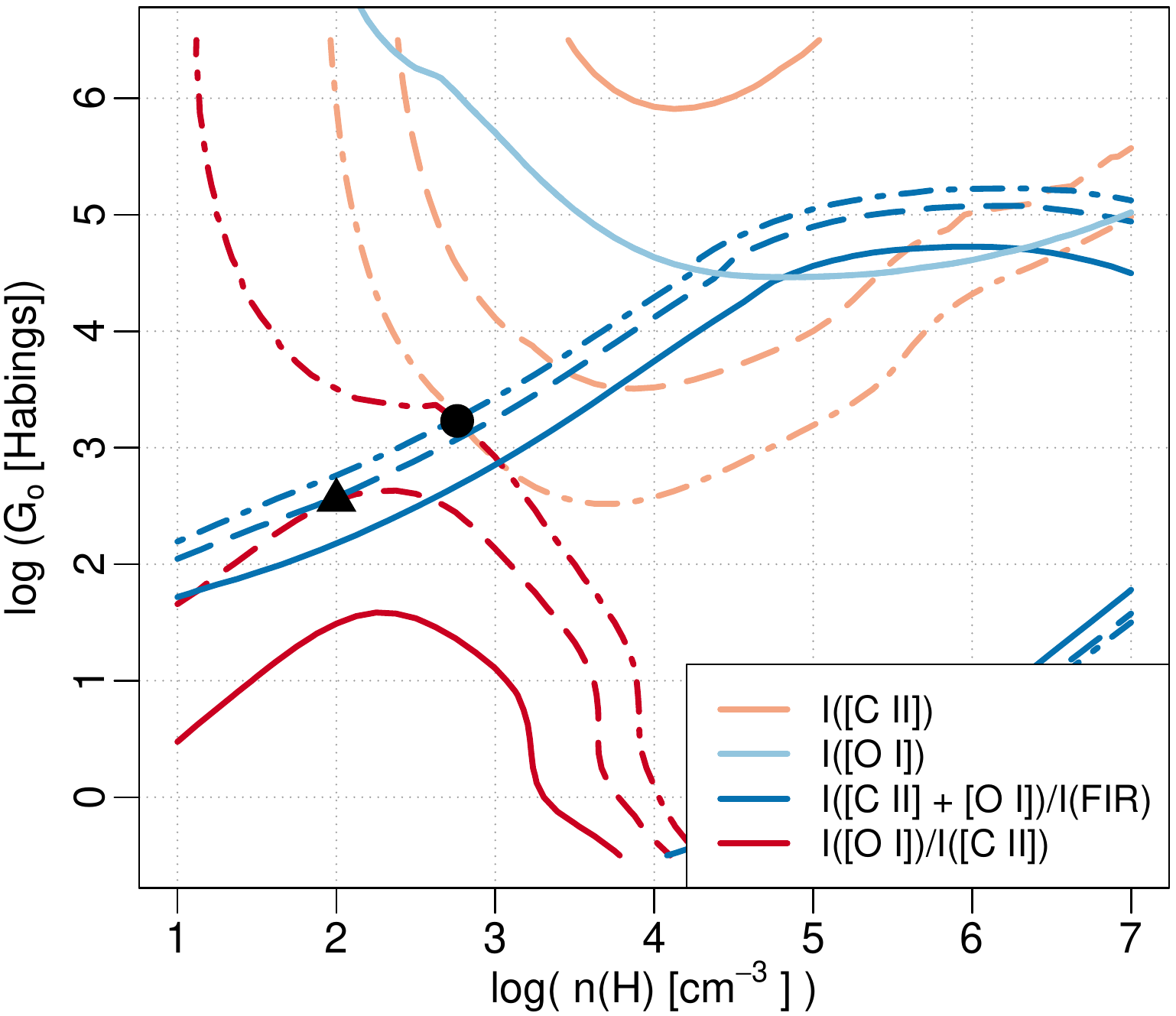}
\caption{ Observed quantities on theoretical PDR results
from PhotoDissociation Region Toolbox models.  Lines in the plot are
contours that correspond to the specific intensities or ratios we
measure or derive in a value vs.\ radiation field strength vs.\
particle density contour plot. (See \citealp{kaufman99} for individual
examples and physical explanations.)  The unmodified observed
quantities (solid lines) are inconsistent with the models.  Model
solutions are possible where lines cross, bounded where 58\% (dashed
lines, triangle) and 76\% (dash-dot lines, circle) of the \cii\
intensity is assigned to \hii\ regions.
\label{fig:pdrgrid}}
\end{figure}

Figure~\ref{fig:pdrgrid} shows our data as contours along predictions
of models from the updated PhotoDissociation Region Toolbox
\citep{pound08, kaufman06, kaufman99}. This model's face-on geometry
matches that of the \sgrb\ region.  Given the changing physical
conditions across \sgrb, we use center-of-error fluxes as guides to
identify boundaries on representative physical conditions, with more
detailed modeling left to future work.  Since \oi\ is only associated
with the $+50$\kms\ clouds, we take only the $+50$\kms\ \cii\ emission
component in the comparison, for $I(\mbox{\cii}) = 1.38\times 10^{-3}$
erg\,cm$^{-2}$\,s$^{-1}$\,sr$^{-1}$.  We use the typical intensity
ratio $I(\mbox{\oi})/I(\mbox{\cii}) = 0.3$ from
Table~\ref{tab:oiIntens} and \citet{goi04}. $I({\rm FIR})$ has been
derived above.  This set of parameters does not produce a model
solution (the solid curves for ratios in the figure do not touch).

Using the total observed \cii\ flux for modeling includes radiation
from \hii\ regions as well as PDRs, however.  Finding successful PDR
model solutions with reduced \cii\ flux allows us to deduce the
relative amounts of \cii\ from PDRs and \hii\ regions.  Ratioed
quantities in the models set a lower limit of 58\% of the \cii\ flux
from \hii\ regions (red and blue dashed ratio lines in
Fig.~\ref{fig:pdrgrid} first touch, marked by a triangle) for
$n(H) \approx 10^{2}$\percmcu\ and a radiation field of
G$_o \approx 10^{2.5}$\,Habing units.  Matching the absolute \cii\ and
\oi\ intensities in this case would require multiple optically thin
PDRs along the line of sight (intensities add, but ratios remain
approximately constant).  Bringing ratios and the \cii\ intensity
together for a single PDR (dash-dot lines cross, marked by dot) sets
an approximate upper limit of 76\% of \cii\ from \hii\ regions of
$n_{\rm H} = 10^{2.8}$\percmcu\ and G$_o = 10^{3.2}$\,Habing units.
These densities and fields are in agreement with those deduced from a
different approach for assessing the \hii\ region contribution, but a
very similar PDR model, using the \cii\ and upper-state 145\um\ \oi\
lines \citet{simpson21}.  Since the line intensity falls rapidly with
decreasing density below the critical density of about
$10^{3.5}$\percmcu, bright emission regions are very likely from
material with density above or near this.  Neither solution matches
the intensity in the ground-state 63\um\ \oi\ line, a frequent problem
between observation and theory for \oi, although the discrepancy could
be reduced by geometries not considered in the model.  Still, with the
predicted $I(\mbox{\oi}) \gtrsim I(\mbox{\cii})$, \oi\ is a major gas
coolant in the PDRs, even as \cii\ remains the dominant coolant for
the region as a whole.

The smaller bright regions identified in Fig.~\ref{fig:ciioispect} and
Table~\ref{tab:oiIntens} require even more \cii\ from \hii\ regions
and have an even larger discrepancy with $I$(\oi).  Many of these
small bright regions could be structures seen edge-on that would not
fit the model's geometrical assumptions, and may have higher densities
as well.

Taking all information together, we find that approximately 50\% of
the \cii\ emission is from \hii\ regions, with the other half from
PDRs.  The proton column density comparison from \cii\ and 20\,cm
radio continuum intensities is more precise; PDR modeling indicates
that this is reasonable, with bounds of about 58\% to 76\%.  These
fractions agree well with values from \hii\ regions obtained from
ratios of \nii/\cii\ intensities in samples across the Galaxy, in the
Arches region of the Galactic center, in the Carina nebula, and
averaged over the nuclei of the galaxies NGC\,891 and IC\,342
\citep{goldsmith15, garcia21, oberst06, stacey10b, roellig16}.
Accounting for \cii\ emission from both PDRs and \hii\ regions is
necessary when interpreting \cii\ emission from galactic nuclei.

\subsection{Physical structure of the \sgrb\ region \label{ssec:physstruct}}

Figure~\ref{fig:sketch} provided an introductory overview of our
understanding of the \sgrb\ region; here we summarize and then explore
the information that led us to those conclusions.  \cii\ emission
stretching from $0.45 \lesssim \ell \lesssim 0.70$ is continuous in
brightness and velocity (Sec.~\ref{ssec:ciispect};
Figs.~\ref{fig:sgrbregion} and \ref{fig:pv}), implying that the entire
region we imaged is physically connected.  The \cii-bright area
encompasses \sgrbo\ and \gsix\ as brighter regions, and at least some
of the source in the \sgrbt\ region.  As we discuss below, the
luminous star formation cores \sgrbt(N), (M), and (S) are essentially
invisible in \cii, and are likely somewhat physically separate from
the extended region, although not entirely so.

\cii\ velocities closely match the velocity field seen in the
H110$\alpha$ hydrogen recombination line, associating \cii\ with the
\hii\ region or regions seen at radio and in mid-infrared spectral
lines \citep{mehringer92, goi04, simpson18b}. At the same time,
brighter \cii\ regions are coincident with \oi\ 63 and 145\um\
emission peaks (Sec.~\ref{ssec:oi}; Fig.~\ref{fig:ciioispect};
\citealt{simpson21}), implying that these regions are edge-on PDRs,
consistent with some exciting stars being mixed with the \cii-emitting
material.  The absence of large-scale intensity gradients in \cii\ or
70\um\ brightness also indicates that the exciting sources are
distributed, rather than being concentrated in one or a few compact
sources.  A similar case can be made at 70\um: despite \sgrbt(M)'s
luminosity, it contributes only 5\% of the total 70\um\ intensity
across the larger \sgrb\ region, and it is not at the center of a
large-scale intensity gradient.

The general velocity field traced along the \cii\ and \hii\ surface
also matches that of the molecular cloud along the same lines of
sight, linking the \cii\ to molecular clouds as well as \hii\ regions
(Fig.~\ref{fig:ciico}, \citealt{mehringer92}).  Molecular and dust
emission is more extended than the \cii-bright region in Galactic
longitude, and even more to increasing longitude
(Fig.~\ref{fig:Integ}).  Lack of \cii\ self-absorption features
indicates that the region bright in \cii\ and \hii\ is on the near
(Earth) side of this cloud or clouds. The velocity field becomes
increasingly complex with increasing longitude beyond \gsix.

\begin{figure}[!ht]
\centering
\includegraphics[width=0.45\textwidth]{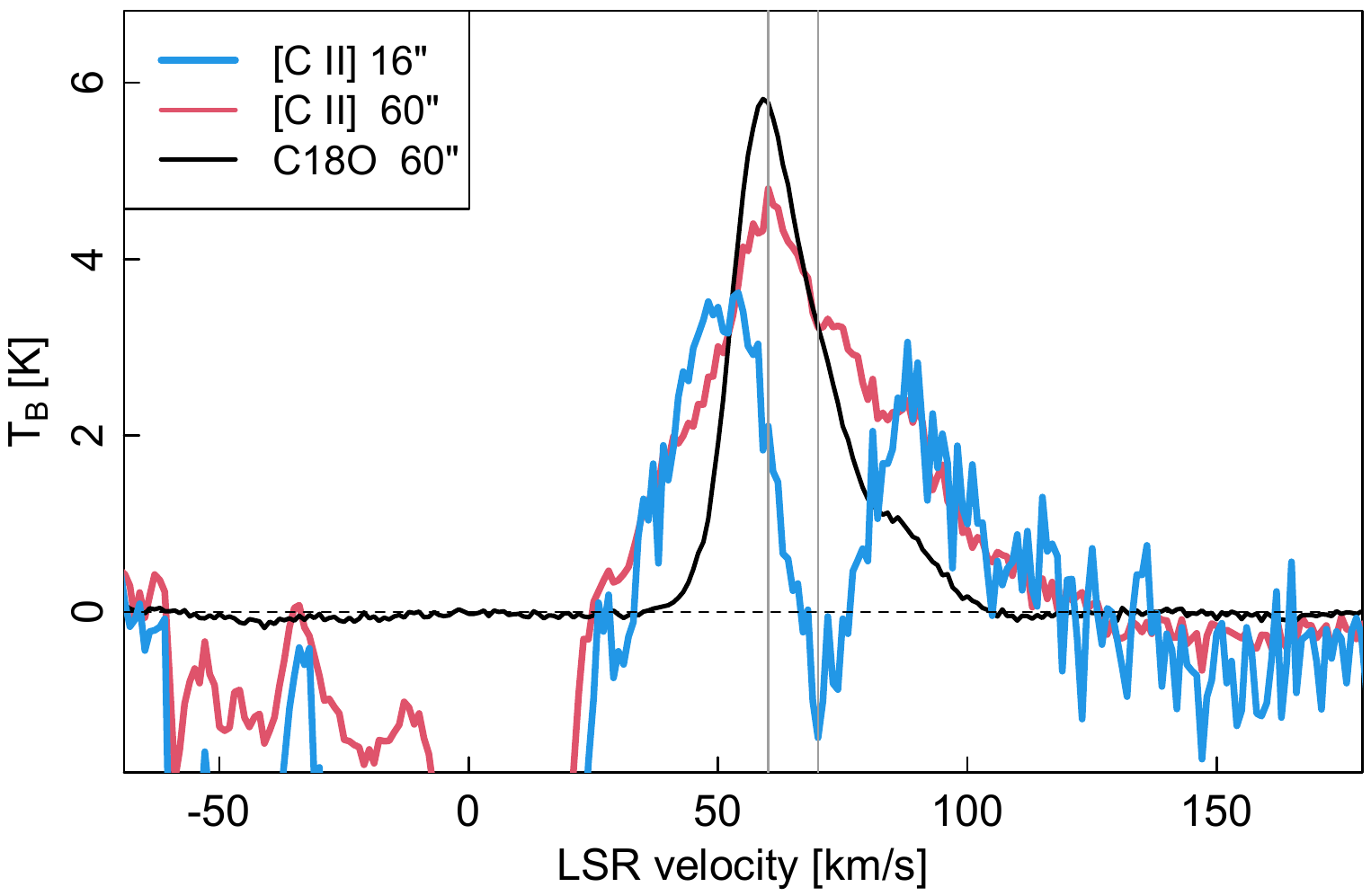} 
\caption{ \cii\ and \ceio\ $J=2-1$ spectra toward \sgrbt(M), showing
the absorption velocity relative to the typical region's velocity
structure.  The overlay compares \cii\ emission averaged over in
16\asec\ and 60\asec\ circular regions and \ceio\ $J=2-1$ over
60\asec. All spectra are on the same brightness temperature scale.
Vertical lines indicate 60 and 70\kms.
 \label{fig:sgrbtmspect}}
\end{figure}

Of all the cores, only \sgrbt(M) has an absorption feature other than
those from the Galactic plane. Absorption is from a source smaller
than the beam that is coincident with the \sgrbt(M) continuum peak.
Figure~\ref{fig:sgrbtmspect} shows this 70\kms\ feature in a 16\asec\
FWHM beam (blue line, duplicated spectrum J of
Fig.~\ref{fig:spectralstack}).  Both visual comparison of the spectra
in Fig.~\ref{fig:sgrbtmspect} and the two-component Gaussian fit shows
that the absorption is at a different velocity than the emission peak.
From Table~\ref{tab:fits}, the absorption is centered at
$70.0\pm 0.3$\kms, while the emission component is centered at
$67.3\pm 0.2$\kms.  The absorption is shifted somewhat from the
65\kms\ cm-wave H$_2$CO and \hi\ absorptions \citep{mehringer95,
  lang10}, although \citet{qin08} find multiple velocities in their
SMA imaging of submillimeter H$_2$CO absorption lines, with the
strongest absorptions at 68 and 76\kms.  Considering velocity
centroids, it seems most probable that the \cii\ absorption is
associated with a compact \hii\ region within the \sgrbt(M) complex.
\citet{depree96} found that their sub-source B within the \sgrbt(M)
complex, one of the two brightest sources at 1.3\,cm, has its
H66$\alpha$ emission velocity peak at 71\kms\ (34\kms\ FWHM).  Both
line center and width are in good agreement with the \cii\ absorption
parameters.

The \cii\ absorption dips close to the continuum level, either because
the \cii\ is thermalized at the dust temperature, or more likely
because surrounding emission in the beam's wings adds flux at the
absorption velocity to diminish a deeper absorption.  This latter
effect is apparent in Fig.~\ref{fig:sgrbtmspect}, where the absorption
feature has completely disappeared in the 30\asec\ beam.

The difference in \cii\ emission and absorption linewidths firmly
place \sgrbt(M) on the Earth side of \sgrb's extended \cii\ emission.
If the \sgrbt(M) continuum source were behind the extended \sgrb\
cloud, then the \cii\ absorption would cover the velocity range
typical of that cloud (as it does for the Galactic plane features),
but the absorption is considerably narrower.  Instead, \sgrbt(M) is
most likely embedded in the dark lane seen against \sgrb's extended
\cii\ emission (see Fig.~\ref{fig:ciiDust}).  Similar cm-wave
absorption line center velocities against the \sgrbt\ cores' continua
(e.g., \citealt{mehringer93, lang10, mills18}) and spatial association
of the three \sgrbt\ cores and other star formation along the dark
dust lane \citep{ginsburg18} suggest that the molecular absorption
velocity is from the high column of material that forms the dark lane,
and that all of the cores are within the lane.  The dark dust lane may
well be the physical manifestation of the ``moderate density
envelope'' \citet{huettemeister95} and others suggest surrounds the
star-forming cores.

A lack of obvious \cii\ or \hii\ emission from the lane's outer
regions suggests that it is either some distance from the stars
irradiating the \sgrb\ molecular cloud behind it, or that the dust has
sufficient optical depth to absorb most of the \cii\ emitted by the
side of the lane between the \sgrb\ stars and the Earth.  Some UV must
be present to produce the \cp\ causing \sgrbt(M)'s 70\kms\ absorption
feature, but since absorption features are missing toward \sgrbt(N) or
(S), the material must be local to structure within \sgrb(M), as
suggested by the \citet{depree96} H66$\alpha$ velocity measurements.

Given the close association between \sgrbt(M) in velocity and
projected distance to other star formation across the overall \sgrb\
region, it seems likely that the \sgrbt\ cores and dust lane, while
somewhat separate from the larger \sgrb\ emission region, are still
physically related, and are not completely separate objects.

\subsection{``Missing'' \cii\ flux from the \sgrbt\ cores \label{ssec:missing}}

Figure~\ref{fig:sgrbtmspect} compares area-averaged \cii\ emission in
16\asec\ FWHM and arcminute-scale beams centered on \sgrbt(M).  In
emission averaged over 1\amin\ diameter areas, \cii\ and \ceio\ have
single-component fit peak velocities of 64 and 62\kms, with 61\kms\ in
a $25^{\prime\prime} \times 25^{\prime\prime}$ area for H110$\alpha$
\citealt{mehringer93}; the absorption feature is distinctly displaced,
at 70\kms. Central velocity comparisons show that the emission
component of the 16\asec\ spectrum is characteristic of the area
surrounding \sgrbt(M), rather than of the source seen in absorption.
Lineshape similarities away from the absorption feature
in Fig.~\ref{fig:sgrbtmspect} further imply that the core is a minor
contributor to the region's \cii\ luminosity at most.  Indeed, it has
no obviously detectable \cii\ emission.

The insignificant amount of \cii\ emission from \sgrbt(M) and the
other luminous cores is especially striking, since they are regions
with extreme star formation densities.  A low \cii/FIR intensity ratio
compared with that in galactic star formation regions and starburst
galaxies (e.g., Sec.~\ref{ssec:comparisons}, \citealt{stacey91,
  stacey10, delooze14, pineda14, herreracamus15, herreracamus18i}), is
reminiscent of the ``\cp\ deficit'' first discovered in (ultra)
luminous infrared galaxies, or (U)LIRGs, by \citet{fischer99}.  With
an FIR luminosity of at least a ${\rm few} \times 10^6$\lsun (e.g.,
\citealt{odenwald84, lis91, gordon93, schmiedeke16}) and a maximum
linear size of the star formation region of about 1\,pc (e.g.,
\citealt{ginsburg18}), a conservative lower limit on its luminosity
density exceeds a ${\rm few} \times 10^{12}$\lsun\,kpc$^{-2}$.  Even
allowing for substantial beam dilution observations averaging multiple
similar regions spread over large scales in distant galaxies, this is
in the range of beam-averaged luminosity surface brightnesses for
ULIRGs of $> 10^{10}$\,M$_\odot$\,kpc$^{-2}$ (e.g.,
\citealt{herreracamus18i}).

This comparison of \sgrbt(M) with (U)LIRGs is not new (e.g.,
\citealp{goi04, kamenetzky14, santamaria21}); our contribution to the
discussion is the combination of velocity resolution and spatial
dynamic range that definitively shows that the emitted \cii\ from the
cores themselves have negligible intensity, even as the surrounding
area provides \cii\ more typical of extended star forming regions.

A number of explanations have been advanced to explain the low
\cii/FIR ratio with increasing FIR: high optical depth, dust-bounded
\hii\ regions that emit strong FIR but little \cii, intense UV fields
that destroy \cp, grain charging that reduces the number of
photoelectrons exciting PDRs, and others (\citealt{luhman98,
  malhotra01}; see also recent summaries in \citealt{herreracamus18ii}
and \citealt{santamaria21} for more details and references). In their
study of a wide range of IR-luminous galactic nuclei,
\citet{herreracamus18ii} found that the dominant mechanism suppressing
\cii\ in most (U)LIRGs is a reduction in photoelectric heating
efficiency as the ionization parameter increases; they note that
optical depth may be important as well.  Comparisons of \cii\ emission
intensities within and just beyond the edge of the dark lane near
\sgrbt(M) provide estimates of lane attenuation factors of 3--5 for
\cii\ emission.  Density gradients around the cores themselves will
increase the attenuation toward the most luminous regions, so
attenuation alone can quite plausibly reduce any \cii\ intensity from
dense cores by more than an order of magnitude.

To the extent that \sgrb\ is representative of regions around extreme
star formation in galactic nuclei, we must be seeing a mixture of
extended and compact star formation in external galaxies, with
weighting toward extended and relatively unobscured regions, and away
from the densest star formation cores.  Proximity and a wealth of data
at many wavelengths all make \sgrbt(M) a valuable local region to
continue detailed investigations into the physical mechanisms
operating within (U)LIRG star formation regions.

\vspace{10mm}
\subsection{Implications for Galactic center star formation
  models \label{ssec:starformation}}
There is debate about the origin of star formation across \sgrb.
\citet{mehringer92} proposed that \sgrbo's extended ionization comes
from a dissipating cluster of mostly O stars that formed earlier than
the clusters and compact \hii\ regions in \gsix\ and \sgrbt.
\citet{goi04} and \citet{simpson18b} also found that late-O stars are
consistent with most of the amount and degree of ionized material in
the \hii\ region that lies between the \sgrb\ background dust and
molecular cloud and the Earth.

The \sgrbt\ cores' more recent burst of star formation suggests some
kind of trigger, most probably associated with a special position
along the region's orbit.  The existence of the dark lane and its
association with the cores suggests that shock-driven propagating star
formation across \sgrb\ is at play.  Narrow dust lanes are common
signs of shock-concentrated material in other galaxies with nuclear
bars and in interacting galactic nuclei.

Star formation triggered by cloud-cloud collisions near \sgrbt\ had
been suggested by \citet{hasegawa94}, among others over the years,
supported by observations of shock-tracing SiO (e.g.,
\citealt{huettemeister95, armijos20, santamaria21}).  The \cii\ p-v
diagram (Fig.~\ref{fig:pv}) shows that linewidths broaden
substantially with increasing longitude from \gsix, possibly signaling
accelerated gas motions from interactions between the \sgrb\ region
and the material to yet higher $\ell$ traced in CO and 160\um\
emission (Fig.~\ref{fig:Integ}).  \citet{simpson18b} also concluded
that shocks or other mechanisms are needed to produce the very
energetic radiation needed to explain the high excitation species they
observe in the \sgrb\ \hii\ region.  An alternative to shocks that
they propose, a tidal tail of hot stars from the dissolving Quintuplet
cluster, as \citet{habibi14}'s modeling might suggest, seems a less
attractive explanation than shocks because it requires a group of
stars with just the right geometry to excite the \sgrb\ area of the
molecular cloud.

There are two popular dynamical models for triggered star formation in
gas orbiting the Galactic center: One proposes that gas cloud
compression by tidal interactions at orbital pericenter (e.g.,
\citealt{kruijssen15, barnes17}) dominates.  The other proposes
cloud-cloud collisions close to the apocenters (e.g.,
\citealt{binney91, sormani20, tress20}) as the chief mechanism.

The first model accounts for the spatial differences in star formation
between the extended star formation in \sgrbo\ and the highly
concentrated clusters of the \sgrbt\ cores by proposing that the two
regions could be on very different positions along their orbits.  In
this model, \sgrbt\ is on the near side of its orbit around the center
after a relatively recent encounter with the central mass
concentration that compressed it and triggered star formation.
\sgrbo, on a similar orbit, passed the center well before \sgrbt, and
has now passed its apocenter and is returning toward the center on the
far side of its orbit.  Its stars have disrupted their birthplaces and
are drifting apart, accounting for the more extended region.  The
positional alignment between \sgrbo\ and the \sgrbt\ cores is by
chance.  As \citet{simpson1819} and \citet{simpson18b} point out,
however, one problem with this explanation is that the difference in
stellar ages between \sgrbo\ and \sgrbt\ is larger than orbital
timescales predict.  If the \sgrbt\ cores are indeed associated with
the dark lane, another problem is in explaining how this long, narrow
structure would be produced and then persist over a substantial
fraction of its orbit around the Galactic center.  Typical velocity
widths in molecular absorptions features toward the \sgrbt\ cores
imply a dark lane lifetime well under $10^6$\,yr, considerably shorter
than the orbital time since periapse.

In the second model, \sgrbo\ and \sgrbt\ are physically related, and
close to an apocenter of their orbit.  Cloud-cloud collisions between
individual gas clouds or streams flowing from the outer galaxy into
the central region along $x_1$ orbits \citep{x1x2} and clouds on the
$x_2$ orbits around the center \citep{sormani20, tress20}, or
interactions between gas on cusped or crossing orbits
\citep{jenkins94} of the $x_2$ orbits themselves, then trigger star
formation. Association of vigorous star formation with the dark dust lane,
and the increasingly complex \cii\ velocity field with increasing
latitude across \sgrb\ are qualitatively consistent with this
model. Such a model for \sgrb, which accommodates different stages of
star formation across the associated regions that comprise \sgrb, is
very appealing.  

\section{Summary \label{sec:summary}} 

Our large-scale, fully velocity-resolved spectroscopic imaging of
\cii\ has revealed:

\begin{enumerate}[noitemsep]
  
\item With 12\% of the total \cii\ flux from 6\% of the area of our
  large-scale image covering all of the \cii-bright Central Molecular
  Zone, \sgrb\ is a major contributor to the entire Galactic center's
  \cii\ luminosity.
  
\item The \sgrb\ region extends as a continuous, coherent structure
  that encompasses the luminous components identified in early radio
  continuum maps of the Galactic center.  \sgrb\ starts near $\ell
  \approx 0.44\degr$, contains \sgrbo\ ($\ell \approx 0.48\degr$),
  runs along the Galactic plane past \gsix, and continues behind and
  beyond \sgrbt\ ($\ell \approx 0.66\degr$) to $\ell \gtrsim
  0.72\degr$. This is a span of some 34\,pc in longitude, with width
  of some 15\,pc in latitude.  Fainter emission extends further,
  particularly to lower $\ell$ toward Sgr\,A.  Many of the region's
  components appear to be part of the same physical structure,
  although the \sgrbt\ region appears as a foreground region that may
  or may not have a physical connection to the larger region.
  
\item The spectra of \cii\ in the \sgrb\ region show two main velocity
  components at $\sim$50\kms\ and $\sim$90\kms , with additional
  emission from $-30$ to $+130$\kms.  The dominant emission is
  centered at $\sim$50\kms, in agreement with the region's
  velocity in molecular and radio recombination lines.  The secondary
  component at $\sim$90\kms\ is mainly associated with extended,
  moderately excited, molecular material.  Individual components have
  typical linewidths of 30--60\kms\ FWHM, indicating emission from a
  highly turbulent medium.  A linear gradient of the mean velocity
  along \sgrb's body runs from $\sim$40 to $\sim$70\kms\ in the same
  sense as Galactic rotation.

\item Emission across the \sgrb\ region is spatially complex.  Arcs,
  ridges, and other structures have nearly exact counterparts in
  70\um\ and 20\,cm continua \citep{molinari16, mehringer92, lang10}.
  Other tracers share the same general distribution, but have
  significant differences from \cii\ and each other.  160\um\
  continuum and CO line emission highlight extended emission from 
  extended background clouds, while 21\um\ and 8\um\ continua
  highlight compact sources and structure.

\item The absence of obvious self-absorption in \cii\ spectra, spatial
  agreement with 70\um\ and 20\,cm radio continuum, and disparity with
  the more extended CO and 160\um\ spatial distributions, indicate
  that the \cii\ is emitted from the near surface of a larger
  molecular cloud or cloud complex.

\item Comparison of a variety of tracers indicates that 70\um, \cii,
  and 20\,cm continua are all excellent tracers of the UV flux
  produced by young stars across the \sgrb\ region.  Velocity
  information in \cii\ is invaluable for separating physical
  components along the line of sight.  Longer-wavelength UV tracers
  are especially important for the Galactic center and other edge-on
  observations of galactic nuclei where even infrared obscuration is
  influential.

\item Velocity resolved \cii/\oi\ flux ratios from a sample of bright
  regions are close to those from large scale measurements
  \citep{goi04, simpson21}, suggesting that PDRs are present across
  the entire \sgrb\ region.  The lineshape we measure for the \oisw\
  line is similar to the $+50$\kms\ \cii\ component, indicating that
  it is mainly associated with \sgrb\ itself, rather than other clouds
  along the line of sight.  Its lineshape also indicates that the \oi\
  line is not significantly affected by self-absorption, and can be a
  significant gas coolant.

\item PDR modeling places bounds of approximately 58\% to 76\% of
  \cii\ flux from \hii\ regions toward \sgrb, with \cii\ and radio
  continuum fluxes suggesting a fraction of more than 50\%.  These
  values agree with those obtained with an independent method using
  \nii/\cii\ intensity ratios. Emission from both PDRs and \hii\
  regions are important in interpreting \cii\ emission from galactic
  nuclei.

\item Distributed star formation is common across \sgrb.  The vast
  majority of UV illumination comes from sources other than the
  \sgrbt\ embedded cores, which produce only 5\% of
  the 70\um\ flux across the region.

\item In \cii, the \sgrbt\ cores appear only as enhanced \cii\
  absorption against their 158\um\ \sgrbt(M) continuum.  \sgrbt(M) may
  be an analog of the intense star formation regions in ULIRGs, which
  exhibit a ``\cp\ deficit'' (e.g., \citealt{fischer99}) when compared
  to their star formation rates.  This region is an excellent local
  source to study the phenomenon in detail.

\item The \sgrbt(M), (N), and (S) cores appear to be objects within
  with a dark dust lane in front of the larger \sgrb\ region.
  Velocity and positional coincidence suggest that the cores and lane
  are still dynamically associated with the larger region, even as
  they are somewhat separated from it.

\item Taken together, our results support triggered star formation
  models that invoke local cloud-cloud collisions close to apocenters
  of orbits in a barred potential (e.g., \citealt{binney91,
    sormani20}), rather than models that invoke tidal compression at
  pericenter passage close to a central mass concentration (e.g.,
  \citealt{kruijssen15, barnes17}).
    
\end{enumerate} 

\acknowledgments We thank the many people who have made this joint
U.S.-German project possible, including the SOFIA observatory staff
and Science Mission Operations former directors E.\ Young and H.\
Yorke.  We also thank M.\ Wolfire and M.\ Pound for their insights and
assistance with PDR modeling, and C.\ Lang for access to her 20\,cm
data.  We thank an anonymous referee for suggestions and requests that
substantially improved this paper.

This work is based on observations made with the NASA/DLR
Stratospheric Observatory for Infrared Astronomy (SOFIA).  SOFIA is
jointly operated by the Universities Space Research Association,
Inc. (USRA), under NASA contract NNA17BF53C, and the Deutsches SOFIA
Institut (DSI) under DLR contract 50 OK 0901 to the University of
Stuttgart. Financial support for this work was provided by NASA
through awards 05-0022 and 06-0173 issued by USRA, the
Max-Planck-Institut f\"ur Radioastronomie and the Deutsche
Forschungsgemeinschaft (DFG) through the SFB~956 program award to
MPIfR and the Universit\"at zu K\"oln. R.S.\ acknowledges support by
the French ANR and the German DFG through the project ``GENESIS''
(ANR-16-CE92-0035-01/DFG1591/2-1).

This research made use of {\em Spitzer} data from the NASA/IPAC
Infrared Science Archive, which is operated by the Jet Propulsion
Laboratory, California Institute of Technology, under contract with
the National Aeronautics and Space Administration; from the {\it
  Herschel} Science Archive, which is maintained by ESA at the
European Space Astronomy Centre; and data products from the Midcourse
Space Experiment, whose data processing was funded by the Ballistic
Missile Defense Organization with additional support from NASA Office
of Space Science.

\vspace{3mm} 

\facilities{SOFIA (\grt)} \citep{young12, risacher18} 

\software{GILDAS (\citealt{pety05, 2013ascl.soft05010G}), 
  SAOImageDS9 (\citealt{joye03, 2000ascl.soft03002S}), 
  R (\citealt{r}), 
  PhotoDissociation Region Toolbox (\citealt{pound08, kaufman06}) 
  \&  
  Baseline correction using splines
  (\url{https://github.com/KOSMAsubmm/kosma\_gildas\_dlc}; 
  \citealt{higgins11, kester14, higgins21})
}


\bibliographystyle{aasjournal}  
\bibliography{sgrb}


\end{document}